\documentclass[aps,prx,reprint,showpacs,preprintnumbers,amsmath,amssymb,superscriptaddress,longbibliography]{revtex4-1}
\usepackage{graphicx}% Include figure files
\usepackage{mathpazo, microtype, mhchem, siunitx, booktabs}
\usepackage{subcaption}

\begin{document}

\frenchspacing

\title[Negative thermal expansion in ScF$_3$]{Quantitative understanding of negative thermal expansion in scandium trifluoride from neutron total scattering measurements}

\author{Martin T Dove}
\email{martin.dove@icloud.com}
\affiliation{Schools of Computer Science and Physical Science \& Technology, Sichuan University, Chengdu 610065, People's Republic of China}
\affiliation{Department of Physics, School of Sciences, Wuhan University of Technology, 205 Luoshi Road, Hongshan district, Wuhan, Hubei, 430070, People's Republic of China}
\affiliation{School of Physics and Astronomy, Queen Mary University of London, Mile End Road, London, E1 4NS, UK}

\author{Juan Du}
\affiliation{School of Physics and Astronomy, Queen Mary University of London, Mile End Road, London, E1 4NS, UK}

\author{Zhongsheng Wei}
\affiliation{School of Physics and Astronomy, Queen Mary University of London, Mile End Road, London, E1 4NS, UK}

\author{David A Keen}
\affiliation{ISIS Facility, Rutherford Appleton Laboratory, Harwell Campus, Didcot, Oxfordshire, OX11 0QX, UK}

\author{Matthew G Tucker}
\affiliation{Oak Ridge National Laboratory, Neutron Scattering Division, 1 Bethel Valley Road, Oak Ridge, TN 37831, USA}

\author{Anthony E Phillips}
\affiliation{School of Physics and Astronomy, Queen Mary University of London, Mile End Road, London, E1 4NS, UK}

\begin{abstract}
Negative thermal expansion (NTE)---the phenomenon where some materials shrink rather than expand when heated---is both intriguing and useful, but remains poorly understood. Current understanding hinges on the role of specific vibrational modes, but in fact thermal expansion is a weighted sum of contributions from every possible mode. Here we overcome this difficulty by deriving a real-space model of atomic motion in the prototypical NTE material scandium trifluoride, ScF$_3$, from total neutron scattering data. We show that NTE in this material depends not only on rigid unit modes---the vibrations in which the scandium coordination octahedra remain undistorted---but also on modes that distort these octahedra. Furthermore, in contrast with previous predictions, we show that the quasiharmonic approximation coupled with renormalisation through anharmonic interactions describes this behaviour well. Our results point the way towards a new understanding of how NTE is manifested in real materials.
\end{abstract}
%\pacs{61.50.Ks,61.66.Hq,64.60.Ej,64.70.K-}
                             

%\submitto{\JPCM}

\maketitle

\section{Introduction}

Almost all materials expand when heated, but some shrink instead. This phenomenon of \emph{negative thermal expansion} (NTE) \cite{Dove:2016bv,Chen:2015hn,Lind:2012jwa,Romao:2013ch} is of fundamental interest from a structural and thermodynamic point of view, and also commercially important \cite{Takenaka:2012cv,Barrera:2005dt,Liang:2010uj}, for instance in preparing substrates resistant to thermal shock. It is among the most widely studied of the anomalous negative thermodynamic properties, others including auxetics with negative Poisson's ratio \cite{Ren:2018ie}, and materials which soften under pressure (negative derivative of the bulk modulus with pressure) \cite{Fang:2013gp,Fang:2013fj, Wei:2020}.

%Our current understanding of the most typical type of NTE, namely that arising from vibrational rather than magnetic or electronic reasons, is based on a simple rationalisation involving the behaviour of particular vibrational modes\cite{Dove:2016bv,Barrera:2005dt,Mittal:2018fy}. One simple visualisation is of atoms joined by stiff bonds into a straight-line chain:
%%\begin{figure}[h]
%\begin{center}
%\includegraphics[width=0.45\textwidth]{chain.pdf}
%\end{center}
%%\end{figure}
%Vibrations of the atoms in the direction perpendicular to the chain will necessarily draw the ends closer together. This has become known as the \emph{tension effect}.\cite{Barrera:2005dt,Dove:2016bv} 

\begin{figure}[b]
\begin{center}
\includegraphics[width=0.5\textwidth]{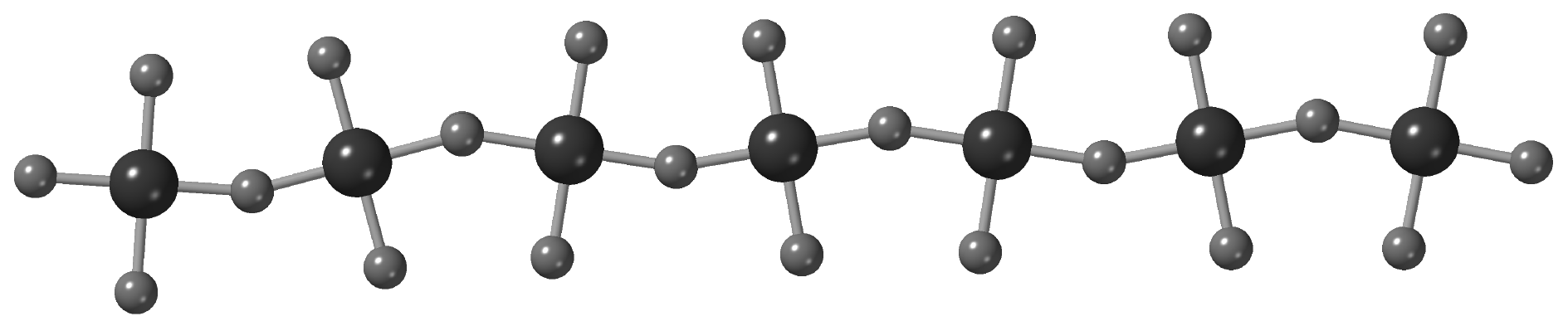}
\caption{Representation of a linear arrangement of corner-linked structural octahedra showing rotations. For clarity the upper atoms are not shown.}
\label{fig:cartoon}
\end{center}
\end{figure}

At the present time we only have a \textit{qualitative} understanding of the general principles underlying the origin of NTE arising from vibrational rather than magnetic or electronic reasons, based on an idea called the `tension effect' \cite{Dove:2016bv,Barrera:2005dt,Mittal:2018fy}. We illustrate this idea in Figure \ref{fig:cartoon} for a linear arrangement of octahedral groups of atoms. Rotations of neighbouring polyhedra will give rise to a transverse displacement of the shared vertex atom. If the bonds between the central and vertex atoms are strong, the transverse displacement of this atom, $u$, will pull its neighbours inwards rather than stretching the bond. If these transverse motions arise from phonons of angular frequency $\omega$, classical harmonic phonon theory gives $\langle u^2 \rangle = k_\mathrm{B}T/m\omega^2$, where $T$ is the temperature. By geometry, if the bonds do not change their length, the thermal motion reduces the lattice parameter $a$ from a value $a_0$ at low temperature to $a \simeq a_0(1 -\langle u^2 \rangle/a_0^2) = a_0(1+\alpha T)$, giving a negative value of the coefficient of linear thermal expansion, $\alpha = a^{-1} \partial a/\partial T = -k_\mathrm{B}/m a_0^2 \omega^2$ \cite{Dove:2016bv}.

This simple picture is far from a good explanation. For one thing, we have to add to this the effects of all other phonons, many of which (including the bond-stretching vibration) will contribute towards \emph{positive} thermal expansion. The tension effect therefore requires that the associated phonons must represent a sufficiently significant number of the total number of phonons.  Furthermore, the fragment of a structure shown in isolation in Figure \ref{fig:cartoon} is part of a three-dimensional crystal structure with the same type of connections in the other two directions. The connections to the rest of the structure give constraints that can significantly reduce the flexibility of the fragment and hence reduce the contribution of the tension effect to thermal expansion. For example, the modulation shown in Figure \ref{fig:cartoon} will require distortions of polyhedra linked in other directions, and in fact in the plane of the diagram there is only one modulation---the one in which neighbouring polyhedra rotate in opposite senses of equal magnitude---that involves no distortions. The energy cost of polyhedral distortions may reduce the effect of such tension-effect vibrations. In view of this discussion, there is currently no physical understanding of why our subject material, ScF$_3$, shows NTE, whereas almost every cubic perovskite material has positive thermal expansion, even though they all have the same basic network structure \footnote{It may be argued that since, in comparison with the perovskite structure, ScF$_3$ has no A-site cation, it should have more flexibility for rotational motions and hence for the tension effect to operate. However, this is a real-space intuition that doesn't necessarily correspond directly with what is really found. Since vibrations are correctly resolved into a summation of normal modes with wave vectors in reciprocal space, any effects of the A-site cation should be interpreted in terms of their effect on the frequencies of the relevant phonon modes. And here we see that the effect is not to block the motion at all. First, comparing for ScF$_3$ \cite{Li:2011dn, Handunkanda:2015dc, Oba:2019bi, Handunkanda:2015dc} and SrTiO$_3$ \cite{Stirling:1972ul} the values of the lowest-frequency modes along the $M$--$R$ directions in reciprocal space, as defined later, we find very similar frequency values and hence the capacity for similar RUM amplitudes. Furthermore, in many cubic perovskites there is a softening of the RUM phonon frequencies on cooling towards a displacive phase transition, which will increase the RUM amplitude.}.
%On the other hand, the coefficient of thermal expansion is a weighted sum over \emph{every} vibrational mode, which in practice is never dominated by a single term. The volume coefficient of thermal expansion $\alpha_V$ of a material is determined by summing the contributions from each vibrational mode, weighted by a factor decreasing with frequency that reflects its thermal accessibility:\cite{Dove:2016bv} 
%\begin{equation}
%  \label{eq:weighted_sum}
%  \alpha_V =\frac{1}{BV}\sum_ic_i\gamma_i.
%\end{equation}
%where $B$ is the bulk modulus, $V$ the volume, $\gamma_i$ the Gr\"uneisen parameter, and the weighting $c_i$ is given by
%\begin{equation}
%  \label{eq:weighting}
%  c_i = \frac{k_\mathrm{B}x_i^2\exp(x_i)}{(\exp(x_i)-1)^2},
%\end{equation}
%where $x_i \equiv \hbar\omega_i/k_\mathrm{B}T$.Thus the phonons that give the tension effect may in fact represent too small a fraction of the total number of vibrations, most of which may contribute to positive thermal expansion.

%Whilst the phonon contributions to NTE have been studied by ab initio calculations, there has been little attempt to study the tension effect in full directly by experiment. A natural way of taking this sum into account is to move to a real-space description of the atomic displacements, which will by definition incorporate the correct contributions from each mode. But it is difficult to develop a realistic atomistic model of NTE directly from experiment, and to our knowledge no such model has previously been reported for any NTE material.

We present here an experimentally-based atomic-scale analysis of NTE in the prototypical material \ce{ScF3} \cite{{Greve:2010bu},{Hu:2016it},{Li:2011dn}}, obtained from neutron total scattering measurements analysed using the Reverse Monte Carlo (RMC) method. This approach is used to refine configurations of atoms so that both their long-range and their local structure are consistent with experimental data. Whilst there have been a few reports of total scattering measurements of NTE materials \cite{Hibble:2002dd,Hibble:2002fv,Tucker:2005hk,Chapman:2005fj,Tucker:2007gk,Dapiaggi:2008et,Chapman:2009fs,Hibble:2013io,Bridges:2014bm,Hu:2016it, Wendt:2019it}---including ScF$_3$ itself \cite{Hu:2016it, Yang:2016ch, Hu:2018eq,Wendt:2019it, Bird:2020dz}---in only one previous case, namely that of ZrW$_2$O$_7$ \cite{Tucker:2005hk, Tucker:2007gk}, has the method been used to generate an atomic model of the fluctuations associated with NTE to provide a consistent examination of the tension effect. From our analysis of the atomic configurations across a wide range of temperatures generated in this study we show that the fluctuations associated with the tension effect are a mix of whole-body rotations and bond-bending distortions of ScF$_6$ octahedra. We have determined the relative balance of these in ScF$_3$ across the range of temperatures in our experiment, and evaluated how this balance leads to NTE in ScF$_3$, whereas similar materials such as SrTiO$_3$ show positive expansivity \footnote{The authors of reference \onlinecite{Li:2011dn} give a slightly misleading qualitative opinion on the fexibility of the ScF$_6$ octahedra on the basis of ab initio molecular dynamics simulations, because they constructed their atomic configuration with an \textit{odd} number of unit cells along each direction. This choice automatically excludes all Rigid Unit Modes, and thus their simulation is unrealistic and their conclusions, albeit qualitative, are affected by this choice.}. The picture that emerges here is consistent across the whole range of temperatures, and supported by simulations using a model system. We also analyse the effects of anharmonicity in ScF$_3$ through the variation of the distribution of atomic displacements with temperature, given some recent calculations of single-model anharmonicity in ScF$_3$ \cite{Li:2011dn}, and our growing understanding that anharmonicity has the effect of reducing NTE at high temperatures \cite{Fang:2014cp, Oba:2019bi}. Total scattering data analysed using the Reverse Monte Carlo method is the only way to obtain information about these issues from experiment.

\section{Background: reciprocal-space model of negative thermal expansion in S\lowercase{c}F$_3$}

\begin{figure}[t]
\begin{center}
\includegraphics[width=6cm]{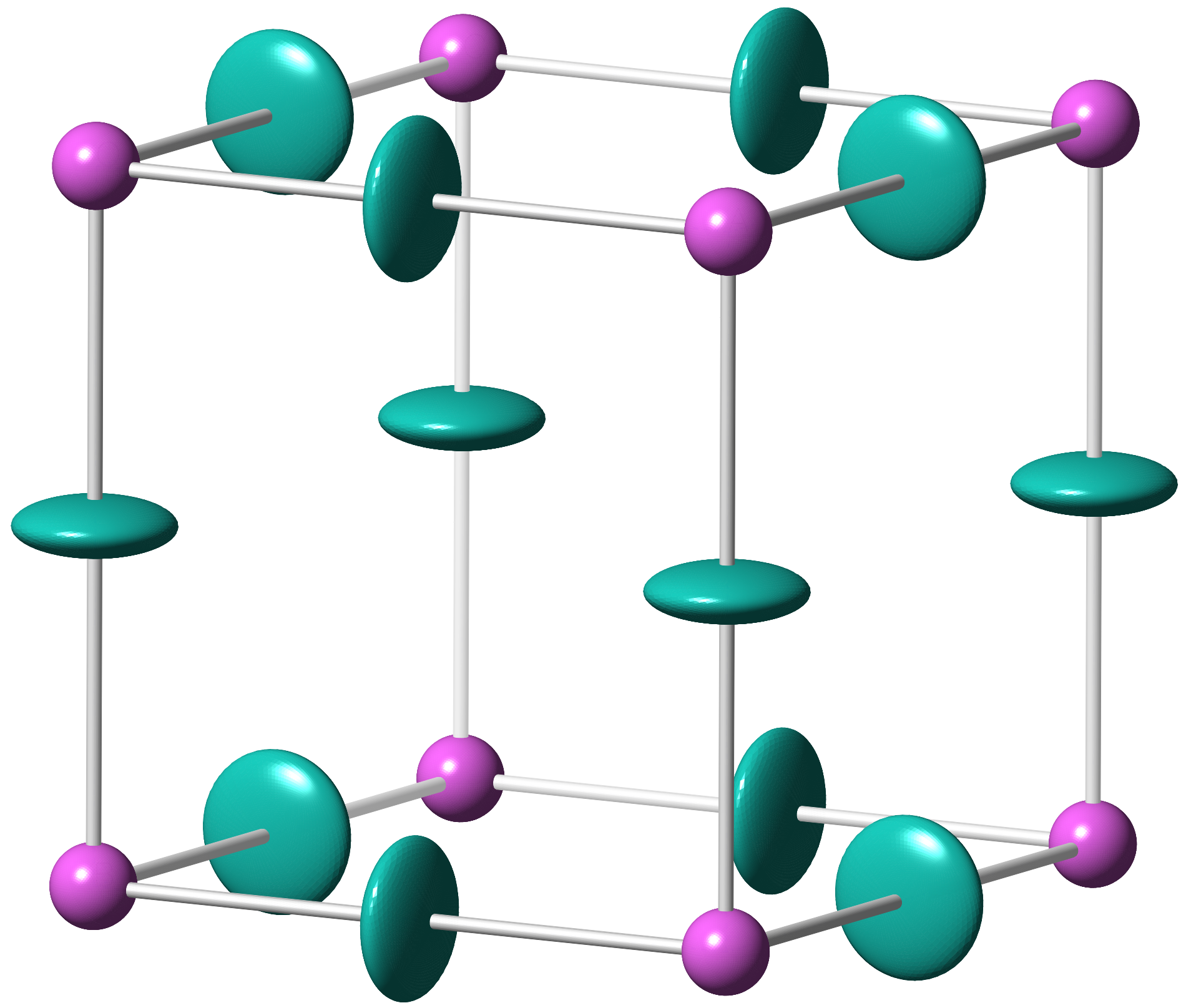}
\caption{Crystal structure of ScF$_3$ at a temperature of 1200 K obtained by Rietveld refinement of neutron powder diffraction data reported in this paper. It has a primitive cubic structure (Strukturbericht symbol $D0_9$, space group $Pm\overline{3}m$) with one formula unit per unit cell). The ellipsoids (Sc pink, F green) represent the thermal motion along different directions, with the volume enclosing 50\% of the total distribution of atom positions.}
\label{fig:crystal_structure}
\end{center}
\end{figure}

Scandium trifluoride, \ce{ScF3}, has the rhenium trioxide structure, equivalent to the perovskite structure with a vacant \textbf{A} site (Figure~\ref{fig:crystal_structure}). It displays isotropic negative thermal expansion over the range \SIrange{0}{1100}{K}---our data are shown in Figure \ref{fig:distances}---with a linear coefficient of thermal expansion of $\alpha = \SI{-10}{MK^{-1}}$ at 200 K \cite{Greve:2010bu}.  

Accurate calculations of the phonon dispersion curves of ScF$_3$ using the Density Functional Theory (DFT) method \cite{Li:2011dn,Handunkanda:2015dc,Oba:2019bi} show two important points. The first is that there is a line in reciprocal space (together with the symmetrically related lines) containing the lowest-energy phonons, namely for the wave vectors  $(\frac12,\frac12,\xi)$ for $-\frac12\leq \xi\leq\frac12$; the two special points $\xi = 0$ and $\xi = \frac 12$ have labels $M$ and $R$ respectively. The eigenvectors of these modes correspond to transverse motions of the F atoms with whole-body rotations of the ScF$_6$ octahedra. Measurements of inelastic x-ray scattering \cite{Handunkanda:2015dc} and diffuse x-ray scattering  \cite{Handunkanda:2016et} from single crystals support this picture exactly. The second point is that the frequencies of phonons away from the $M$--$R$ line increase rapidly with frequency, and these modes have low frequency \textit{only} for wave vectors near this line in reciprocal space. Again, this point is consistent with the inelastic and diffuse scattering measurements  \cite{Handunkanda:2015dc, Handunkanda:2016et}. 

The low-frequency modes with wave vectors along the $M$--$R$ line and with eigenvectors corresponding to octahedral rotations are what are called Rigid Unit Modes (RUMs) \cite{Giddy:1993ue,Hammonds:1996wy,Heine:1999vk, Dove:2016bv, Dove:2019dl}. Their low energy is a consequence of the fact that the force constants associated with bending the octahedral F--Sc--F angles are much larger than those associated with bending the linear Sc--F--Sc angles; our estimate discussed in the Supplemental Material is that the two force constants differ by a factor of around 50. Such a large factor accounts for the fact that the value of the frequency of the transverse acoustic mode at wave vector $(\frac12,0,0)$ is much larger than that of the RUM frequency, seen in both the DFT calculations \cite{Li:2011dn,Handunkanda:2015dc,Oba:2019bi} and inelastic x-ray scattering experiments \cite{Handunkanda:2015dc}. The same situation is seen in the experimental phonon dispersion curves of the cubic perovskite phase of SrTiO$_3$ \cite{Stirling:1972ul}. The existence of RUMs provides a natural mechanism for the tension effect in NTE materials since the rotations give rise to a shrinkage of the crystal structure and these modes have the necessary low energy \cite{Heine:1999vk,Dove:2016bv,Dove:2019dl}. Indeed, the DFT calculations show that the RUMs have a considerably larger contribution to NTE than all other individual phonons, by two orders of magnitude.\footnote{It is worth remarking about the origin of the stiffness in the structural polyhedra in general and more specifically in the case of the ScF$_6$ octahedra in ScF$_3$. Much of the original literature on RUMs concerned silica and silicates, which are conventionally considered to have strong covalent bonds defining the shape and stiffness of the structural SiO$_4$ tetrahedra. However, it is important to understand that the rigidity of structural polyhedra do not rely on covalent bonding, because tension within the polyhedra in a system where the bonding is more ionic in nature can arise from mutual repulsions between the vertex ions because of size effects and Coulomb interactions. This is pertinent for ScF$_3$. The Sc--F bond is relatively strong, as evidenced by the large value of the Sc--F stretching frequency in the phonon dispersion curves \cite{Li:2011dn,Handunkanda:2015dc,Oba:2019bi}, indicating some degree of covalent bonding, but it is more likely that the ionic size and electrostatic interactions will be most important in the F--Sc--F bond-bending forces.}

As we have noted, pure RUMs exist along lines of wave vectors. A line of wave vectors of RUMs in ScF$_3$ occupies only a tiny---effectively infinitesimal---fraction of reciprocal space, and virtually all phonons must necessarily involve distortions of the ScF$_6$ octahedra. In particular, even though the pure RUM motions have the highest contribution to NTE of any phonon, their tiny weighting in reciprocal space means that any tension effect model must involve such distortions. The finite, as opposed to infinite, stiffness associated with distortions of the polyhedra are in fact an important part of the RUM model \cite{Dove:2019dl}; we return to this point in \S\ref{sec:discussion}. And in fact, the DFT phonon calculations \cite{Li:2011dn,Handunkanda:2015dc,Oba:2019bi} show that in reciprocal space the tension effect will `bleed into' the phonons whose wave vectors are close to, but not exactly on, the line $M$--$R$. These modes have a large component of rotation but an increasing component of polyhedral distortion on moving away from the $M$--$R$ line in reciprocal space. They are, in effect, what we would call quasi-RUMs \cite{Hammonds:1996wy}. The extent to which the spectrum of quasi-RUMs can give rise to an overall NTE will depend on the extent to which the polyhedra can easily be distorted, a question that is analysed in detail in this paper. This will give us a new perspective of the role of RUMs in the tension effect that will be applicable to many NTE materials, and will enable us to understand, \textit{for the first time}, why NTE can exist in some materials but not in other materials with close structural similarity.

At this point we note that the authors of reference \onlinecite{Wendt:2019it} have proposed a radically different model. Their central idea is that the ScF$_6$ octahedra have no internal rigidity in terms of bending of the bonds, and that the Sc--F bonds can rotate in an uncorrelated way as independent Einstein oscillators. In part this idea is based on a misleading interpretation of the first three peaks in the pair distribution function because of the use of inappropriate integration limits coupled with the effects of noise associated with the Fourier transforms. The idea of the Sc--F bonds being able to rotate in an uncorrelated motion implies the absence of a force constant associated with bending of the octahedral F--Sc--F bond, which is directly the opposite of the RUM model. This type of model would surely give rise to a significant tension effect. However, it would lead to an excitation spectrum with 6 low-frequency modes for all wave vectors, and an additional two low-frequency shear acoustic modes along the $(\xi,0,0)$ direction and one along the $(\xi,\xi,0)$ direction. Whist this model of uncorrelated Sc--F motion is appealing as an intuitive interpretation of the tension effect, it is completely inconsistent with our knowledge of the phonon dispersion curves by both ab initio calculation  \cite{Li:2011dn,Handunkanda:2015dc,Oba:2019bi} and inelastic x-ray scattering measurements \cite{Handunkanda:2015dc}. A recent paper based on new x-ray total scattering measurements \cite{Bird:2020dz} supports the interpretation presented here, based on an earlier pre-print of this article, over the model of uncorrelated F-atom motions proposed in reference \onlinecite{Wendt:2019it}.

The RUM model with infinite stiffness and the uncorrelated model of reference \onlinecite{Wendt:2019it} represent two opposite ends of a spectrum. We would argue that the calculated \cite{Li:2011dn,Handunkanda:2015dc,Oba:2019bi} and measured  \cite{Handunkanda:2015dc} phonon dispersion curves actually mean that the balance is more towards a RUM model with more polyhedral flexibility than found in corresponding oxides. Where ScF$_3$ actually sits in this balance is explored in detail in this paper.

\section{Methods}
\label{sec:methods}

Neutron total scattering and diffraction experiments were performed on the Polaris diffractometer at the UK ISIS spallation neutron facility. The sample was obtained commercially, and x-ray and neutron powder diffraction measurements showed that the sample is of single phase within the limits of detection. The sample was packed into a cylindrical vanadium can of radius 8~mm. Measurements for 750~$\mu$A.h were obtained over the temperature range 10--1200~K, with shorter runs at intermediate temperatures performed for crystal structure analysis.  The POLARIS instrument can measure down to a wavelength of 0.1~\AA\ \cite{POLARIStech}, which gives a maximum energy transfer far in excess of the upper limit of 85~meV required from the DFT phonon calculations on ScF$_3$ \cite{Li:2011dn} and therefore the experiments and subsequent analysis capture the full range of phonon excitations. 

Rietveld refinement was carried out using the GSAS software  \cite{Larson:2004wv} with the EXPGUI interface \cite{gsasgui}. Data were prepared for Rietveld analysis using the MANTID software \cite{Arnold:2014iy}.

The RMC simulations were performed using the RMC\-profile code \cite{Tucker:2007eh}. The data sets used were the total scattering function after correction and subtraction of the self term, $i(Q)$, the pair distribution function (PDF) $D(r)$ obtained as Fourier transform of the function $Qi(Q)$ (the corrections to form $i(Q)$ and conversion to the PDF $D(r)$ were performed using the GUDRUN package \cite{Soper:2012vs}), and the Bragg scattering profile. Key equations and data are given in the Supplemental Material, showing the high quality of the fitting we were able to achieve. 

Molecular dynamics simulations were performed using the DL\_POLY code \cite{Todorov:2006ee}, using a model developed by fitting the calculated dispersion curves to the DFT results of reference \onlinecite{Li:2011dn} using the GULP lattice simulation code \cite{Gale:1997iq,Gale:2003eo}. The model is described in more detail in Section \ref{sec:MD}, in the Supplemental Material, and in the parallel reference \onlinecite{Wei:2020}.

\section{Real-space analysis of negative thermal expansion}
We collected total neutron scattering data from a powder sample of scandium trifluoride, measuring both the Bragg scattering---sensitive to the long-range order---and the diffuse scattering. Although several previous pair distribution function studies of ScF$_3$ have used X-rays \cite{Hu:2016it, Yang:2016ch,Hu:2018eq, Bird:2020dz}, for our analysis neutron radiation was a more appropriate choice, for three reasons. First, the accessible range of scattering vector $Q$, and hence the resolution of the pair distribution function derived from it, is much greater: we were able to measure up to a maximum value of $Q_\text{max} = 50$ \AA$^{-1}$, while with X-rays the maximum achievable value of $Q$ with a short-wavelength silver anode is $22.5\,\mathrm{A}^{-1}$, and is usually practically up to around 30 \AA$^{-1}$ with synchrotron radiation measurements. Second, the X-ray atomic form factor decays rapidly with scattering vector $Q$, which further limits the $Q$ range in which useful data can be collected: even if we were somehow able to measure X-ray scattering at 50 \AA$^{-1}$, the scattering factor of Sc would be only 0.7\% of its value at low $Q$. Finally, because this $Q$-dependence differs between atoms, calculating a scattering-weighted pair distribution function from a trial configuration of atoms is necessarily approximate.

Thus our neutron data enabled us to calculate high-resolution, bias-free pair distribution functions $D(r)$ (see §\ref{sec:methods}), which are effectively histograms of instantaneous interatomic distances. We then used the Reverse Monte Carlo (RMC) method \cite{{Tucker:2001fh},Dove:2002ed,{Keen:2005dd},{Tucker:2007eh}} to obtain a set of atomic configurations consistent with these data, each of which can be regarded as a plausible snapshot of the instantaneous atomic positions in this material.

%The quality of one of our typical Rietveld refinements is demonstrated in Figure~\ref{fig:rietveld}, showing the fit of the calculated diffraction pattern to the experimental data.

\begin{figure}[t]
\begin{center}
\includegraphics[width=8cm]{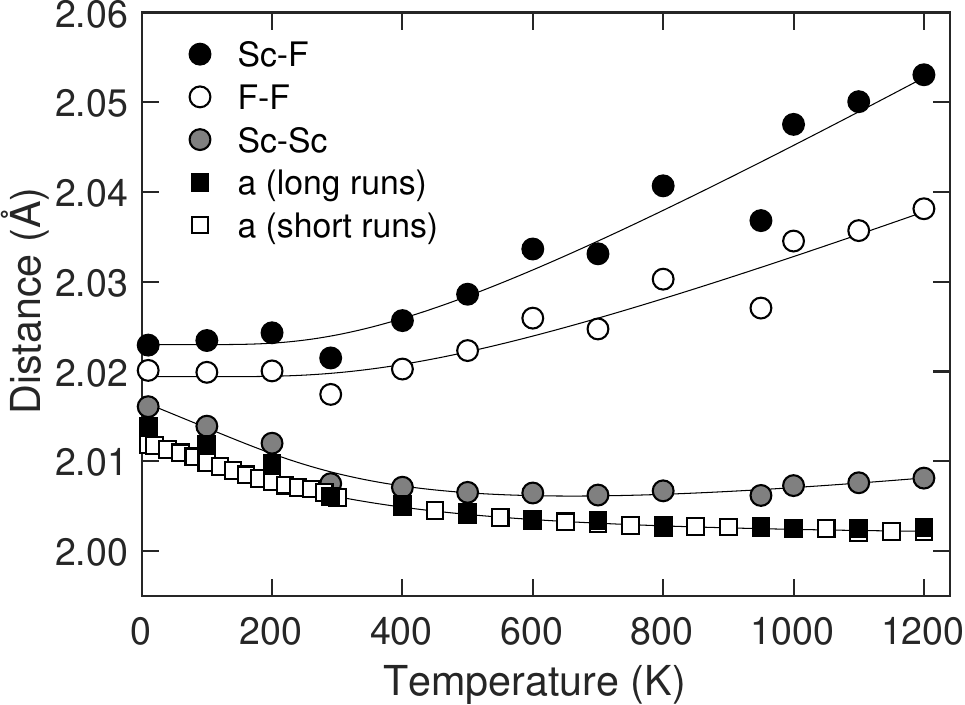}
\caption{Comparison of the temperature dependence of half the lattice parameter, $a/2$ (black squares, filled squares representing data from longer measurements and open squares representing data from shorter measurements), half of the average instantaneous nearest-neighbour \ce{Sc\bond{...}Sc}  distance obtained from analysis of the RMC configurations (gray filled circles), the average instantaneous nearest-neighbour F--F distance obtained from analysis of the RMC configurations scaled by $1/\sqrt{2}$  (open black circles), and the average instantaneous nearest-neighbour Sc--F distance also obtained from analysis of the RMC configurations (black filled circles. In each plot statistical error bars are smaller than the sizes of the data symbols. The scaling means that each data set should converge to a value of $a/2$ at low temperature. The lines are guides to the eye; the guides for the F--F and Sc--F distances were obtained by fitting functions of the form $d = d_0 + \alpha \coth(\theta/T)$, and the guides for the lattice parameter and Sc--Sc distances were obtained by fitting functions of the form $d = d_0 - \gamma T + \alpha \coth(\theta/T)$, where the parameters $d_0$, $\alpha$, $\gamma$ and $\theta$ were variables in the fitting process.}
\label{fig:distances}
\end{center}
\end{figure}

The experimental lattice parameter is shown in Figure~\ref{fig:distances}, plotted as $a/2$, showing the NTE over the temperature range 0--1100~K and positive expansion at higher temperatures consistent with previous data \cite{Greve:2010bu}. In this figure we also compare the mean nearest-neighbour Sc--F and F--F distances, and half the mean \ce{Sc\bond{...}Sc}  distance, all three obtained from analysis of the RMC configurations. The Sc--F and F--F distances are fully consistent with the positions of the peaks in the PDF data, but in the case of the \ce{Sc\bond{...}Sc} distance, the peak in the PDF overlaps with that from the second-neighbour \ce{F\bond{...}F} distribution and thus we cannot extract this directly from the raw data. The prediction from the tension effect is that the distance between mean positions of two bonded atoms should be shorter than the actual mean bond length, and indeed as expected the Sc--F bond and F--F distances show normal positive thermal expansion (Sc--F expansion coefficient $\alpha = +15$~MK$^{-1}$), whereas $a/2$ decreases on heating defining the negative thermal expansion. This result for the Sc--F bond is consistent with two recent measurements of the PDF \cite{Hu:2016it, Wendt:2019it}. \footnote{The recent study of Wendt et al \cite{Wendt:2019it} reports two anomalous findings, first that the mean F--F distance shrinks with increasing temperature, and second that the integrated area of the second F--F peak in the PDF decreases with temperature. In both cases these are contrary to what we have found. These two anomalous results may reflect the limitations of analysis of the PDF alone, without the support of a model. As the F--F peak broadens with temperature, the data become increasingly noisy, and the pair density within any fixed $r$ range will necessarily decrease. Furthermore, the analysis may also be affected by artefacts associated with untreated truncation effects at the maximum observable value of $Q$. Such problems highlight the value of making use of modelling approaches such as RMC, where the atomic model imposes reasonable physical constraints on the data analysis. The atomistic configuration RMC provides means that there is no ambiguity in assigning pair density where peaks are very broad or overlap, such as the 4~\AA\ peak here, which encompasses both Sc--Sc and F$\cdots$F pairs. Figures S7--S9 in the Supplemental Material show the RMC-derived decomposition of the PDF into contributions from individual atomic pairs.}

%It is interesting to observe that the rate of change of the Sc--F bond with temperature ... value of the expansivity of the Sc--F bond increases at higher temperatures, and the point at which the expansivity reaches its maximum value on heating coincides with the temperature at which the crystal thermal expansion turns from negative to positive.

%The comparison of the results for the Sc--F distance with $a/2$ show the important point that the Sc--F distance has positive thermal expansion even though the material shows overall NTE,

%Note that that the distance between the mean positions of nearest-neighbour Sc atoms defines the value of the lattice parameter, and the distance between the mean positions of neighbouring Sc and F atoms is half the value of the lattice parameter. The Bragg scattering measures the mean positions of atoms, and the total scattering measures the mean instantaneous interatomic distances, which are qualitatively and quantitatively different quantities. 

Figure~\ref{fig:distances} also shows the temperature dependence of the mean distance between neighbouring Sc atoms. One might expect, given the locations of the Sc atoms, for this distance to reflect exactly the lattice parameter. However, although slight, we see a difference between these two quantities that grows on heating, with the \ce{Sc\bond{...}Sc} distance showing a slightly weaker dependence on temperature than for the linear dimensions of the crystal, and a change to positive expansivity at a lower temperature. A similar effect was seen in Zn(CN)$_2$, where the (negative) expansivity of the \ce{Zn\bond{...}Zn} distance is less negative than the linear expansivity of the crystal \cite{Chapman:2005fj}. In that case the difference is due to the fact that the primary mechanism for NTE is from the acoustic modes \cite{Fang:2013ji}. This is of course different to ScF$_3$, where the main NTE modes are rotational modes of optic character that lie along the edges of the Brillouin zone. However, by symmetry these modes transform into transverse acoustic modes as the wave vectors changes from the $M$--$R$ line to zero, as seen in the dispersion curves reported in references \onlinecite{Li:2011dn} and \onlinecite{Aleksandrov:2002ut}. We propose that the behaviour of the \ce{Sc\bond{....}Sc} distance may be associated with the growing acoustic mode character of the NTE phonons moving away from the $M$--$R$ line; indeed, the bending of the F--Sc--F right angle can be associated with the transverse acoustic (shear) mode. %Although the contribution to the NTE from an individual mode with a wave vector in the vicinity of, but not along, the $M$--$R$ line is much lower than a mode actually on the line, as seen in the results of reference \onlinecite{Li:2011dn}, these modes occupy a much larger fraction of phase space and therefore will contribute significantly because of this.

\section{Local structural distortions from Reverse Monte Carlo analysis}

\begin{figure*}[t]
\begin{center}
\begin{subfigure}[b]{0.49\textwidth}
\includegraphics[width=0.9\textwidth]{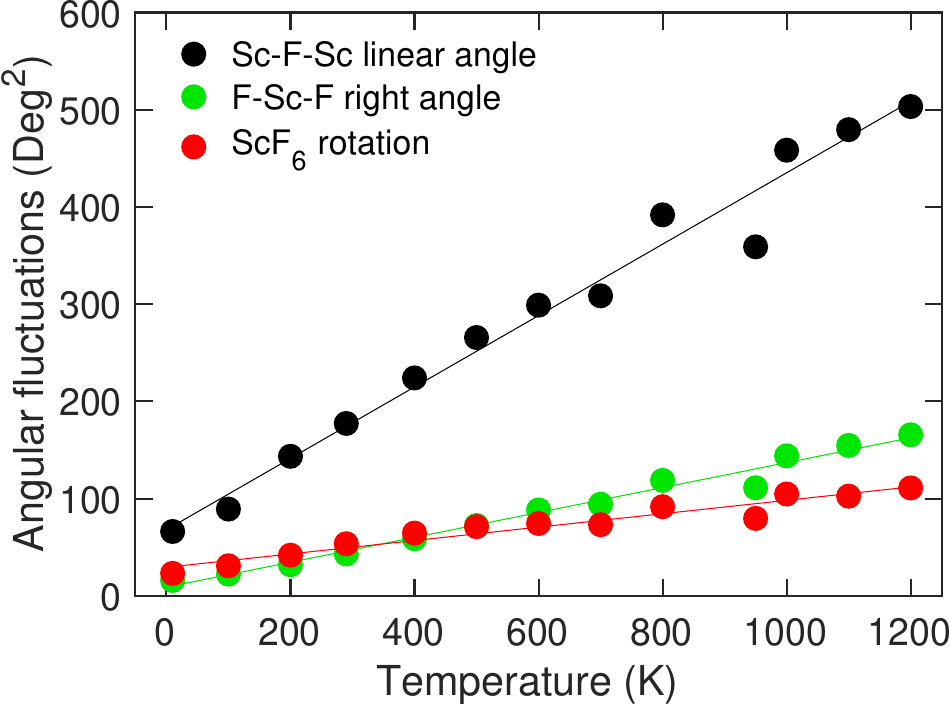}
\caption{}
\label{fig:angles}
\end{subfigure}
\begin{subfigure}[b]{0.49\textwidth}
\includegraphics[width=0.9\textwidth]{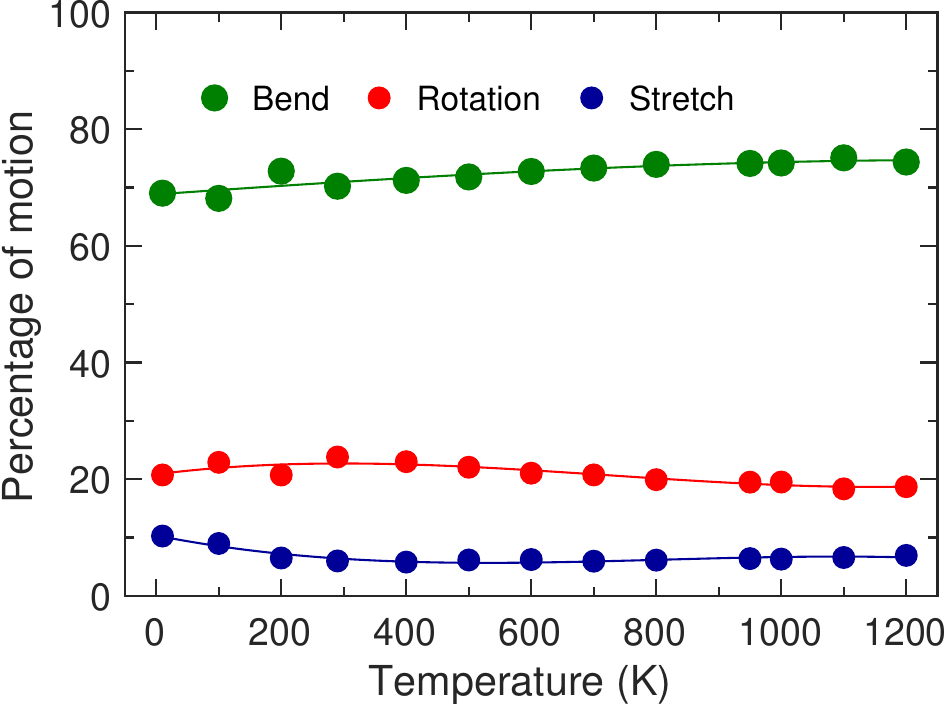}
\caption{}
\label{fig:gasp}       
\end{subfigure}
%\raggedright a) \\
%\centering \includegraphics[width=7cm]{ScF3_devangles.pdf}  \\
%\raggedright b) \\
%\centering  \includegraphics[width=7cm]{ScF3_gasp2.pdf}
\caption{(a) Comparison of the variances of three angles associated with local motions taken from the RMC configurations. The black points show the departure of the Sc--F--Sc angle from its nominal value of $180^\circ$, the green points show the variance of the F--Sc--F angle as it fluctuates from its nominal value of $90^\circ$, and the red points show the mean-square angle of rotation of the ScF$_6$ octahedra as a whole body. (b) Breakdown of the total motion (excluding overall polyhedral displacements) of the atoms in the ScF$_3$ crystal from the GASP analysis of the RMC configurations, where the red, green and blue points and guides to the eye represent the fraction of the motion that is associated with whole-body rotations of the ScF$_6$ octahedra, deformations of the F--Sc--F right angles, and stretching of the Sc--F bonds respectively. In both plots statistical error bars are smaller than the sizes of the data symbols, and the lines/curves are given as guides to the eye. }
\label{fig:angles_and_gasp}
\end{center}
\end{figure*}

We now consider the local atomic motions that are associated with NTE. The fact that the Sc--F bond shows positive thermal expansion implies that the tension effect will provide the mechanism. Thus we need to consider effects associated with transverse motions of the F atom, %due to swinging of the Sc--F bonds, 
and the extent to which this is correlated with the ScF$_6$ octahedra moving as nearly-rigid objects or distorting. Figure \ref{fig:angles} shows the behaviour of three angles with temperature: first the variance of Sc--F--Sc angles as they distort from the value of $180^\circ$, second the variance of the F--Sc--F angles as they distort from the ideal octahedral angle of $90^\circ$, and third the mean-square rotations of the ScF$_6$ octahedra (calculated using the GASP tool; see below). The largest fluctuation, by a significant margin, is for the Sc--F--Sc angle, which is primarily associated with the transverse motions of the F anion and is consistent with both the thermal ellipsoids seen in Figure \ref{fig:crystal_structure} and the role of the tension effect. The other two angles, namely of the ScF$_6$ octahedral rotations and the bending of the F--Sc--F bond, are actually very similar to each other. Thus the transverse motions of the F atoms as reflected in the Sc--F--Sc angles are achieved by both rotations and bond-bending deformations of the ScF$_6$ octahedra.

In Figure \ref{fig:gasp} we show the details of an analysis performed using the GASP method, based on using geometric algebra to represent the rotations of polyhedral groups of atoms \cite{Wells:2002ty,{Wells:2002tq},{Wells:2004en}}. Given a set of bond vectors for an octahedron, $\mathbf{r}_i$, where $i$ runs over all the centroid-vertex bonds, we can compare the vectors in one configuration with those in another (here, the ideal average structure), which we denote as $\mathbf{r}'$. The difference, which we call the \textit{mismatch}, is $\mathbf{e}_i = \mathbf{r}_i - \mathbf{r}'_i$. GASP uses a least-squares algorithm to find the rotation of each octahedron that minimises $M = \sum_i |\mathbf{e}_i|^2$, where we sum over all bonds in the polyhedron.
The residual value of $M$ per polyhedron is then decomposed into contributions from bending of bond angles and stretching of bonds, thereby accounting for the total motion involving non-uniform displacements of the F atoms. The results in Figure \ref{fig:gasp} compare the extent to which the sum of the atomic motions of the F atoms in each ScF$_6$ octahedron can be separated into whole-body rotations of the octahedron, flexing of the F--Sc--F $90^\circ$ bond angle, and stretching/shrinking of the Sc--F bonds. %The key result is that around 70\% of the motions of the F atoms are associated with flexing of the F--Sc--F angles, 20\% with whole body rotations of the octahedra, and 10\% with flexing of the Sc--F bonds.
\footnote{In Figure \ref{fig:angles} the sizes of the fluctuations of the ScF$_6$ orientations and F--Sc--F angles are similar, but in Figure \ref{fig:gasp} the GASP analysis suggest that the motions of the F atoms associated with bond bending are rather larger than from rotations of the octahedra. These results are actually consistent with each other: the contribution to $M$ from the octahedral rotation will come from three axes, but there are 12 bending angles associated with deformation of the octahedra.} 
This partition, which barely changes with temperature, is compared in summary form with corresponding results from a similar study of the TiO$_6$ octahedra in the perovskite SrTiO$_3$ \cite{Hui:2005bg} in Table~\ref{tab:gasp}, together with results from a molecular-dynamics simulation on a model system (described in the Supplemental Material) in which the energy penalty for bending the \ce{X-M-X} angle tends to zero. Our results show that \ce{ScF3} is quite close to that limiting case \footnote{But it would be wrong to say that it \textit{is} the limiting case, because the model, being a simplified model, is not an exact representation of the actual material. In particular, the division between the three types of motion is not exactly the same. What we are demonstrating is that even in the limit of fairly flexible octahedra the GASP method will still identify whole-body rotations.}. The analysis suggests, therefore, that the ScF$_6$ octahedra in ScF$_3$ are significantly more flexible with regard to bending the anion--cation--anion angles than are the TiO$_6$ octahedra in SrTiO$_3$; we will argue below that this is the key difference that gives rise to NTE in \ce{ScF3} but not in the cubic oxide perovskites.  \footnote{There are two important points to highlight in the two diagrams in Figure \ref{fig:angles_and_gasp}. The first is that the angular fluctuations in (a) increase linearly with temperature, in line with expectations from classical quasi-harmonic lattice dynamics. Second is that the distribution of atomic motion over the three components in (b) is more-or-less constant with temperature, reflecting phonon eigenvectors and again being fully consistent with classical quasi-harmonic lattice dynamics. In particular, although thermal motion is large at the higher temperature, it does not lead to any breakdown of the classical picture, contrary to the opinion expressed in reference \onlinecite{Wendt:2019it}.}

\begin{table}[t]
\caption{The percentage mismatch between different atomic configurations of a network of \ce{MX6} octahedra and the ideal structure, decomposed by GASP into \ce{X-M-X} bending, \ce{M-X} stretching, and \ce{MX6} rotation components. We compare three systems: the RMC configurations of \ce{ScF3}, a hypothetical perovskite structure in the limit where the octahedra have flexible bond angles, and \ce{SrTiO3} as also analysed by RMC and taken from reference \onlinecite{Hui:2005bg}. The hypothetical structure is an important comparison because, even if there are no rigidity constraints applied to the bond angles within the ScF$_3$ octahedra, some fraction of their random distortion will always be mathematically attributable to a rigid-body rotation.}
\label{tab:gasp}
\begin{tabular}{lrrr}
\toprule
Material & Bend & Stretch & Rotation \\
\midrule
\ce{ScF3} & 70\% & 10\% & 20\% \\
Flexible model & 75\% & 5\% & 20\% \\
\ce{SrTiO3} & 44\% & 19\% & 37\% \\
\bottomrule
\end{tabular}
\end{table}

%\begin{figure}[t]
%\begin{center}
%\includegraphics[width=8cm]{ScF3_gasp2.pdf}
%\caption[]{Breakdown of the total motion (excluding displacements) of the atoms in the ScF$_3$ crystal, where the blue, magenta and red points and guides to the eye represent the fraction of the motion that is associated with whole-body rotations of the ScF$_6$ octahedra, deformations of the F--Sc--F right angles, and stretching of the Sc--F bonds respectively.}
%\label{fig:gasp}       
%\end{center}
% \end{figure}

Comparing absolute values of the fluctuations for ScF$_3$ and SrTiO$_3$ at a single temperature, say 300 K, we find that in SrTiO$_3$ the linear Ti--O--Ti angle fluctuates by an average of around $5^\circ$ and the TiO$_6$ octahedra orientation fluctuates by around $2^\circ$  \cite{Hui:2005bg}, while the corresponding sizes of the fluctuations in ScF$_3$ are around $14^\circ$ and $7^\circ$ respectively (Figure \ref{fig:angles}). On the other hand, the coefficient of thermal expansion of the Ti--O bond, 10~MK$^{-1}$ is comparable to that of the Sc--F bond cited above, with similar Sc--F and Ti--O bond lengths.

We can also compare this analysis with the phonon dispersion curves for ScF$_3$ calculated using ab initio methods \cite{Li:2011dn} and calculated for SrTiO$_3$ using a shell model fitted to inelastic neutron scattering and infrared spectroscopy \cite{Stirling:1972ul}. We see that the octahedral cation--anion stretching frequencies are very similar (20 THz in ScF$_3$ and 22 THz in SrTiO$_3$), suggesting (given the similar masses of O and F) that the bonds in ScF$_3$ and SrTiO$_3$ are of similar stiffness. In both ScF$_3$ and SrTiO$_3$ the octahedral rigid-body rotational phonons between the $R$ and $M$ wave vectors are of very low frequency compared to the stretching mode in both materials, namely between 0.6 to 1.2 THz in ScF$_3$ and between 1.3 to 2.5 THz in SrTiO$_3$ at a temperature of 200 K. However, the bending frequencies are different. If we take, for example, the transverse acoustic mode frequency at $X$, $(\frac{1}{2},0,0)$, which is a shear mode that reflects bending distortions of the octahedra, we find that it is lower in frequency by around a factor of 2 in ScF$_3$ than in SrTiO$_3$, meaning the corresponding force constants are 4 times smaller. This is consistent with our finding that the ScF$_6$ octahedra are rather more flexible than the TiO$_6$ octahedra. But care is needed not to go to the extreme viewpoint and imagine that there is no force constant associated with the bending of the F--Sc--F right angle. There is, and it directly gives rise to a non-zero shear elastic constant \footnote{We consider models in which the Sc--F bond is the sole rigid entity, with complete flexibility of the ScF$_6$ octahedra, as for example in reference \onlinecite{Wendt:2019it}, to be over-flexible and unable to provide useful insights, not least because such models will predict the wrong signs of the Gr{\"u}neisen parameters for many phonon modes.  A similar situation was discussed with regard to NTE in Cu$_2$O in reference \onlinecite{Rimmer:2014dj}, where the true rigid entity is the linear O--Cu--O trimer, and models that consider only the O--Cu bonds as the rigid entities are too flexible.}.

\section{Molecular dynamics simulations with a simplified model}\label{sec:MD}

\begin{figure*}[t]
\begin{center}
%\raggedright a) \\
%\centering \includegraphics[width=6cm]{Gaussians1.pdf} \\
%\raggedright b) \\
%\centering \includegraphics[width=6cm]{Gaussians2.pdf} \\
%\raggedright c) \\
%\centering \includegraphics[width=6cm]{PeakWidth.pdf}
\begin{subfigure}[b]{0.45\textwidth}
\includegraphics[width=1\textwidth]{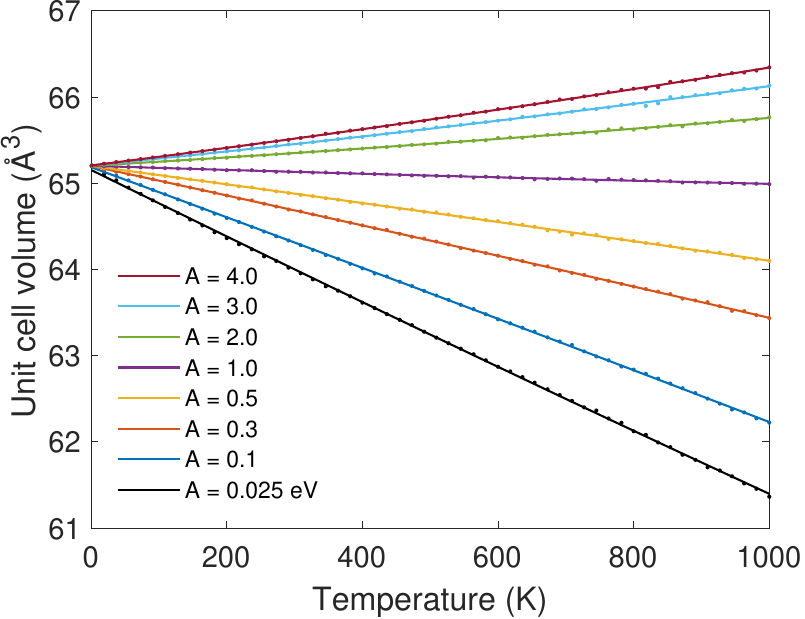}
\caption{}
\end{subfigure}
\begin{subfigure}[b]{0.45\textwidth}
\includegraphics[width=1\textwidth]{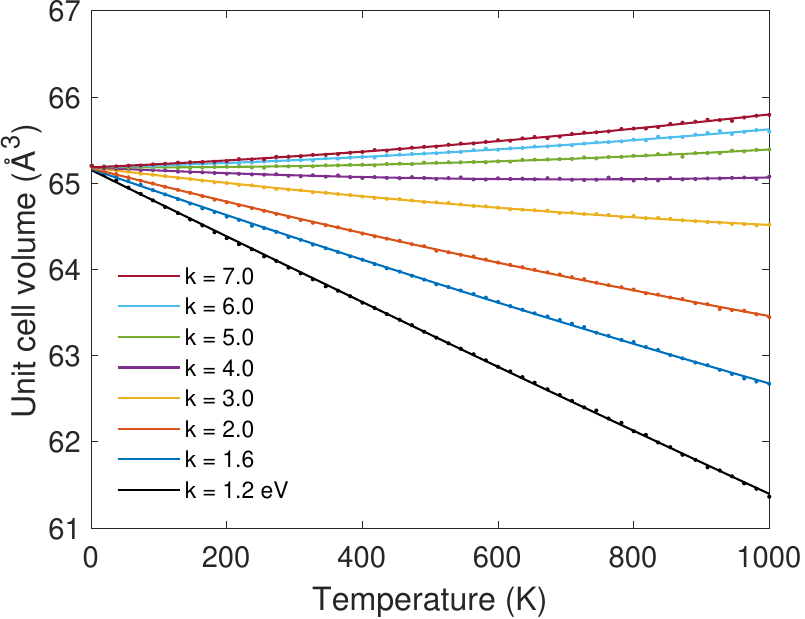}
\caption{}
\end{subfigure} \\
\begin{subfigure}[b]{0.45\textwidth}
\includegraphics[width=1\textwidth]{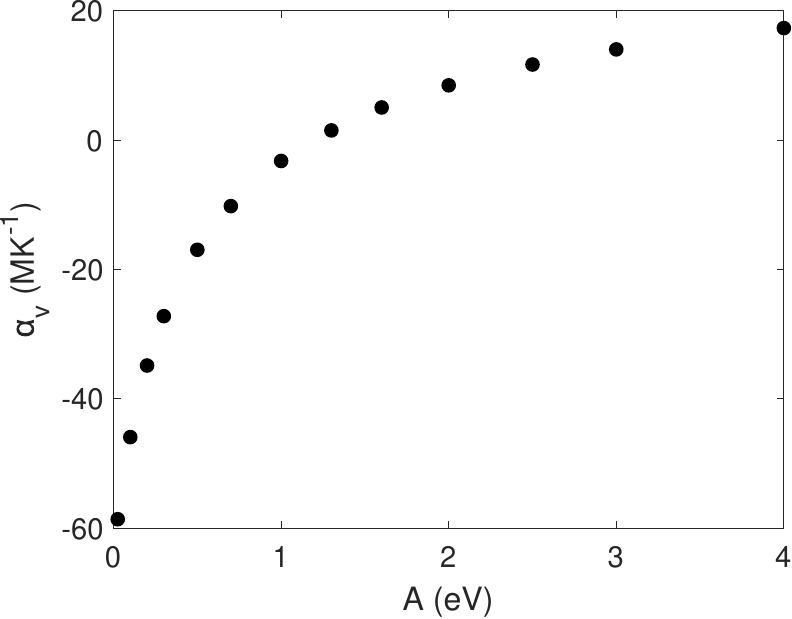}
\caption{}
\end{subfigure}
\begin{subfigure}[b]{0.45\textwidth}
\includegraphics[width=1\textwidth]{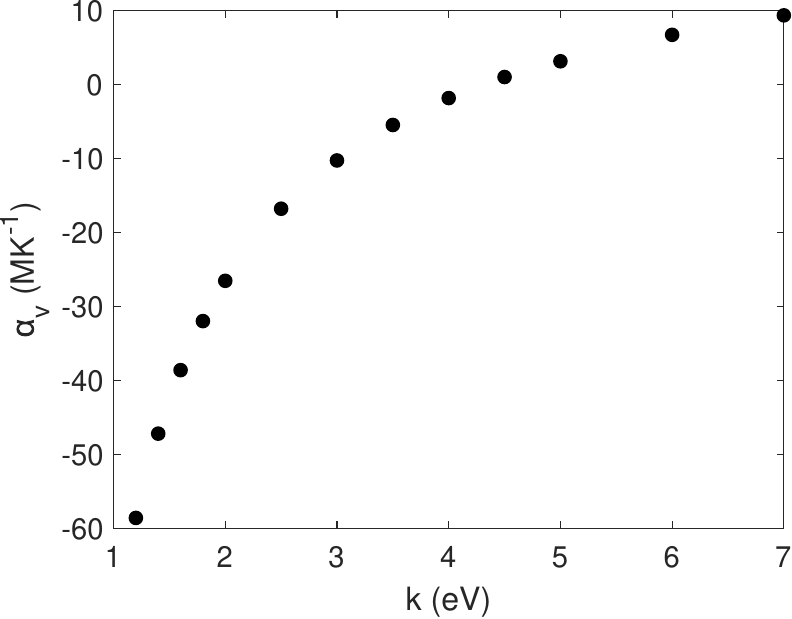}
\caption{}
\end{subfigure}
\caption[]{(a) and (b) show the variation of volume of one unit cell of ScF$_3$ with the inter-octahedral Sc--F--Sc and intra-octahedral F--Sc--F angle force constants respectively, $A$ and $k$ (see Supplemental Material for the equations). (c) and (d) show the corresponding coefficients of volumetric thermal expansivity for the cases $k = 1.2$ eV and $A = 0.025$ eV respectively, corresponding to the values that give best match to the DFT dispersion curves \cite{Li:2011dn} as discussed in the Supplemental Material.}
\label{fig:MDresults}       
\end{center}
\end{figure*}

Simulation methods can often give insights into the relationship between the properties of materials and their atomic structure. For ScF$_3$ there have been a number of simulation studies using the molecular dynamics (MD) method with with classical force fields or ab initio methods \cite{Lazar:2015es, Bocharov:2019kj, Bocharov:2020iu}. What is less useful about such methods is that it is not easy to change parameters that directly affect one type of structure flexibility alone. Any change in some aspect of the model will affect everything. To address the question of the relative roles of the forces associated with the bending of the octahedral F--Sc--F bond angle or linear Sc--F--Sc bond angle we need to work with a simpler idealised model, and we explore this now.

The model introduced briefly in the previous section, and in our parallel paper on pressure-induced softening in ScF$_3$ \cite{Wei:2020}, is described in more detail in the Supplemental Material. The model has been designed and analysed to see the effect of various independent parameters on the NTE. There are two parameters that are of interest (a third parameter controls the bond stretching, which is tuned to a high stiffness by comparison with the DFT phonon calculations). The first, with symbol $A$, controls flexing of the linear Sc--F--Sc bond and determines the frequency of the RUM along the $M$--$R$ wave vectors. The value of this parameter was tuned directly by comparison of the calculated RUM frequencies with those given by the DFT phonon calculations. The second, with symbol $k$, controls flexing of the F--Sc--F right-angle, and its value was tuned to reflect the variation of the shear acoustic modes from the DFT dispersion curves. Increasing both of these parameters will reduce the flexibility of the structure in their respective ways, and hence change the thermal expansion. We explicitly do not include ionic charges in the model because Coulomb interactions will affect the flexibilities of both the  linear Sc--F--Sc bond and right-angle F--Sc--F bond. Nevertheless, the simple three-parameter model does a surprisingly good job of reproducing the phonon dispersion curves, comparing Figure S11 with the results, say, of reference \onlinecite{Li:2011dn}.

Figure \ref{fig:MDresults} shows the results of varying both force constants starting from the model that best reproduces the DFT phonon dispersion curves. The first result from the data shown in Figure \ref{fig:MDresults} is that increasing both force constants will reduce the negative thermal expansion, and eventually drive it positive. Increasing the F--Sc--F force constant $k$ will increase the frequencies of the modes close to the RUM $M$--$R$ line, which will reduce the number of phonons contributing significantly to the overall NTE and hence leading to a reduction and eventual elimination of NTE. This is consistent with the narrative developed based on the RMC results presented above. On the other hand, the force constant associated with the linear Sc--F--Sc bond, $A$, plays a role in increasing the frequencies of the RUMs but less of a role in shaping the dispersion curves, so will have a gradual effect in changing the NTE as the mode frequencies increase. The increased frequency of the RUMs will lead to a reduced transverse amplitude of the F atoms. In some ways, this is similar to the effects of the forces imparted in the analogous perovskites with an atom in the \textbf{A} site (such as Sr in SrTiO$_3$). There is, however, one significant difference with regard to changing the two force constants. In the case of the F--Sc--F right angle, NTE vanishes by increasing the force constant value by a factor of 3.3, whereas in the case of the Sc--F--Sc angle, NTE vanishes by increasing the force constant value by around a factor of 40. Looking at dispersion curves in perovskites, for example in SrTiO$_3$ \cite{Stirling:1972ul}, suggests that in perovskites the size of this factor is not reached. On the other hand, the factor of 3.3 increase in the bending force constant is exactly consistent with the difference between ScF$_3$ and SrTiO$_3$ discussed in the previous section. Thus the difference strongly suggests that key effect in determining NTE is indeed the bond-bending flexibility of the ScF$_3$ octahedra.

\begin{figure*}[t]
\begin{center}
%\raggedright a) \\
%\centering \includegraphics[width=6cm]{Gaussians1.pdf} \\
%\raggedright b) \\
%\centering \includegraphics[width=6cm]{Gaussians2.pdf} \\
%\raggedright c) \\
%\centering \includegraphics[width=6cm]{PeakWidth.pdf}
\begin{subfigure}[b]{0.32\textwidth}
\includegraphics[width=0.98\textwidth]{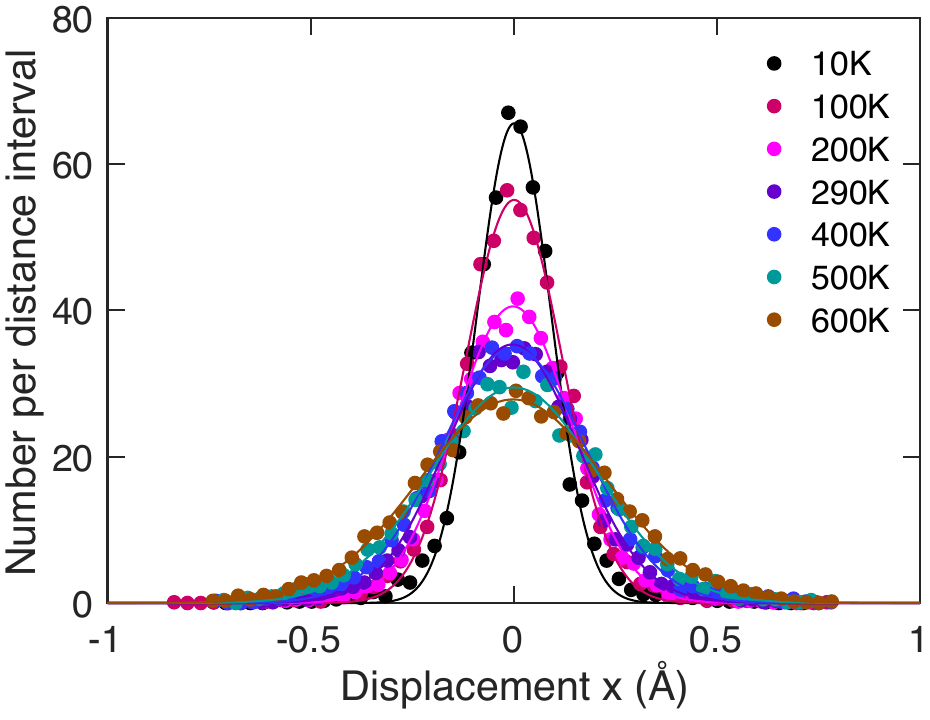}
\caption{}
\end{subfigure}
\begin{subfigure}[b]{0.32\textwidth}
\includegraphics[width=0.98\textwidth]{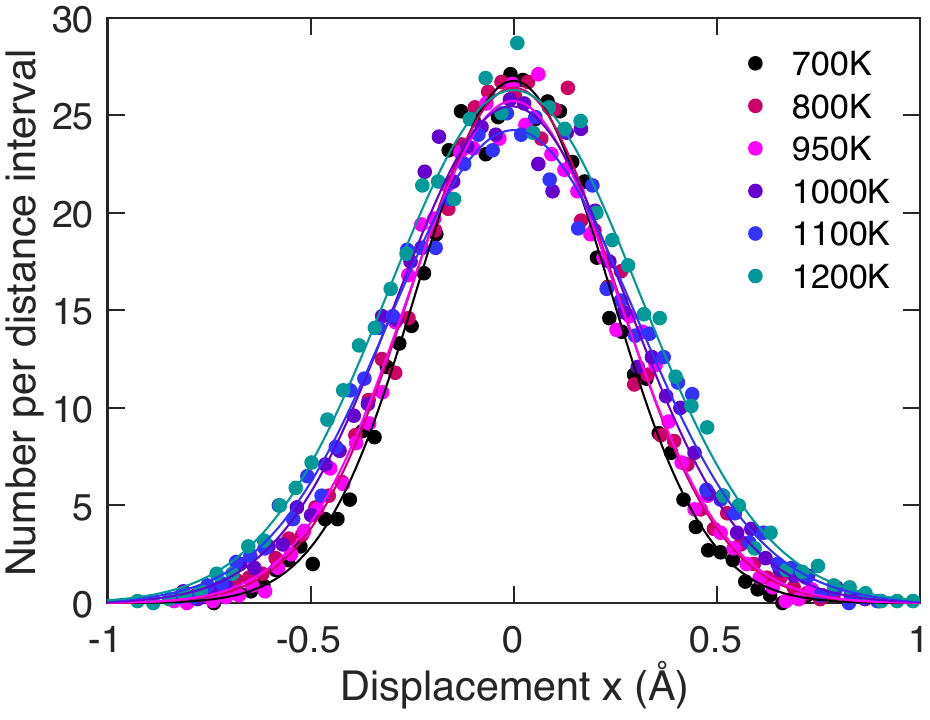}
\caption{}
\end{subfigure}
\begin{subfigure}[b]{0.32\textwidth}
\includegraphics[width=0.98\textwidth]{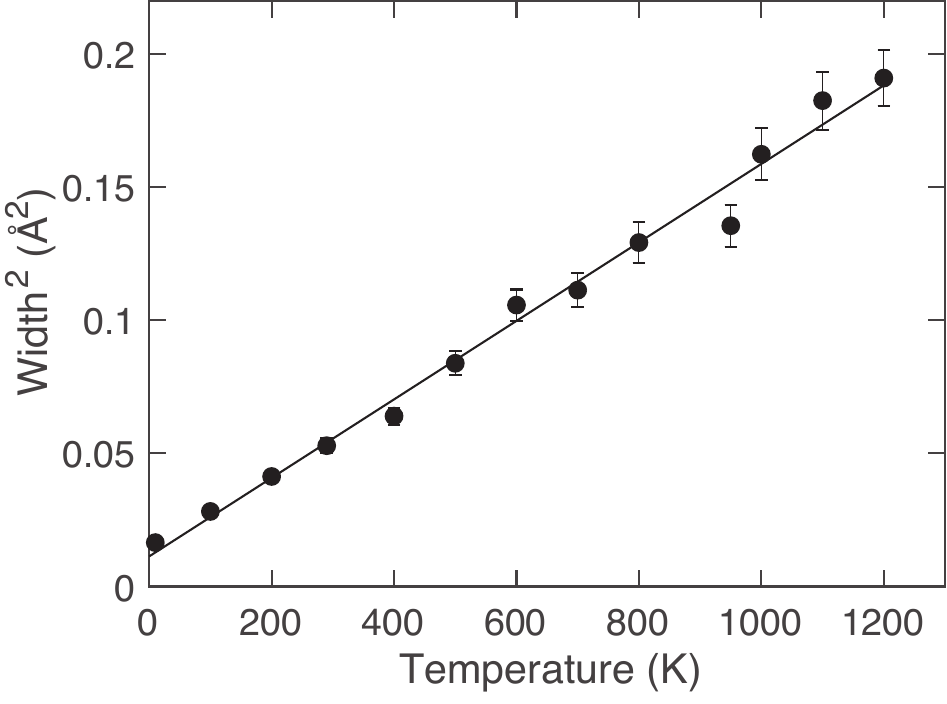}
\caption{}
\end{subfigure}
\caption[]{(a) and (b) show histograms of the lateral displacements of the F atoms at the different temperatures (circles). At all temperatures the distributions are well described by Gaussian functions (thin lines). (c) shows the variance of the Gaussian fits as a function of temperature. The data follow a linear dependence on temperature throughout the range studied, as indicated by the fitted straight line, demonstrating that a renormalised phonon model is sufficient to describe the anharmonic effects in \ce{ScF3}.}
\label{fig:gaussians}       
\end{center}
\end{figure*}

\section{Anharmonicity}
There is a lot of current interest in the role of anharmonic phonon interactions in NTE. Typically the most important ones are those involving fourth-order interactions, which have the effect of changing phonon frequencies. Several recent papers have studied anharmonicity in ScF$_3$ in various ways  \cite{Li:2011dn,Handunkanda:2015dc,vanRoekeghem:2016kd,Oba:2019bi}. %In a recent paper we showed how such anharmonic effects will affect the NTE, shifting in the direction of positive expansion at higher temperatures as a result of the NTE phonons increasing in frequency on heating. In the case of ScF$_3$ the NTE phonons have wave vectors around the line M--R in reciprocal space. Inelastic x-ray scattering measurements have shown that their frequencies vary strongly with temperature similar to the behaviour of a classical soft mode associated with a phase transition, as seen in SrTiO$_3$ for example.

In renormalised phonon theory \footnote{In common usage, the term \emph{quasiharmonic} model considers the effect of changes in volume on the phonon frequencies through the consequent changes in harmonic force constants, giving new harmonic phonon frequencies as modified by the anharmonic coupling of force constants to volume. The term \emph{renormalised phonon} model considers how a renormalised harmonic Hamiltonian can result from a mean-field treatment of the anharmonic interactions between phonons, typically focussing on the fourth-order interactions.} the temperature-dependence of a phonon angular frequency $\omega(\mathbf{k},j)$ ($j$ labels the mode for any $\mathbf{k}$)  in the high-$T$ limit varies as \cite{Dove:2009bn,Oba:2019bi}
\begin{equation} \label{eq:rnpt}
\omega^2(\mathbf{k},j) = \omega_0^2(\mathbf{k},j) + \tfrac{1}{2} k_\mathrm{B}T \sum_{\mathbf{k}^\prime,j^\prime} 
\frac{\alpha_4(\mathbf{k},\mathbf{k}^\prime,j,j^\prime)}{\omega_0^2(\mathbf{k}^\prime,j^\prime)} \nonumber
\end{equation}
where $ \omega_0^2$ is the square of the harmonic angular frequency, and the interactions characterised by the fourth-order anharmonic parameters $\alpha_4$ couple the phonon $(\pm \mathbf{k},j)$ to all other phonons $(\pm \mathbf{k}^\prime,j^\prime)$. This summation includes the case $(\mathbf{k},j) = (-\mathbf{k}^\prime,j^\prime)$ ; when this case is taken alone, it represents the \emph{independent-mode approximation}. It is this approximation \textit{only} that is probed in a frozen-phonon calculation \cite{Li:2011dn}, and it will give just a small part of the overall picture. That is, the contributions from the modes $(\mathbf{k},j) \ne (\mathbf{k}^\prime,j^\prime)$ are normally much more important than only the modes $(\mathbf{k},j) = (-\mathbf{k}^\prime,j^\prime)$ in determining how phonon frequencies change with temperature. The DFT calculations of Li et al \cite{Li:2011dn} suggested that for the $R$-point mode the independent-mode anharmonic potential is quite large compared to the harmonic potential, but the summation over all modes may still mean that the primary anharmonic effects come instead from interactions across the Brillouin zone. Van Roekeghem et al \cite{vanRoekeghem:2016kd} recently studied the anharmonicity using both x-ray inelastic scattering and through calculations of the phonon frequencies via a renormalised phonon method. They showed, consistent with most perovskites, that the low frequency branch along the line $M$--$R$ and the three lowest-frequency optic modes at zero wave vector soften on cooling, consistent with renormalised phonon theory, whereas the higher-frequency modes harden on cooling. Similar results were obtained by Oba et al \cite{Oba:2019bi}. The softening on cooling arises from direct anharmonic interactions via renormalised phonon theory as described here, whereas the hardening of the high-frequency modes arises primarily from thermal expansion of the Sc--F bond. In this model, the renormalised phonons continue to look like phonons with well-defined frequencies, with lifetimes substantially larger than the phonon frequency (See Figure 3 of reference \onlinecite{vanRoekeghem:2016kd}). Separately, in reference \onlinecite{Fang:2014cp} we showed from simple considerations that anharmonic renormalisation of phonon frequencies will cause NTE to shift towards positive expansivity at higher temperatures; the same result is obtained by more detailed renormalised phonon theory calculations \cite{vanRoekeghem:2016kd,Oba:2019bi}.

We have analysed our RMC configurations to look for any effects of anharmonicity in the distributions of transverse displacements of fluorine atoms. This particular atomic displacement was chosen since it is active in the $R$-point RUM previously identified as having a dominant fourth-order term. Figure~\ref{fig:gaussians} shows the distribution of these displacements of F atoms away from the \ce{Sc\bond{...}Sc} line as a function of temperature. Two features of these data are noteworthy. First, the distributions are well fitted by Gaussian functions at all temperatures. In particular, we find no evidence for a toroidal distribution of fluorine atoms, as conjectured in a recent PDF study \cite{Wendt:2019it}. \footnote{Importantly, the reverse Monte Carlo method is indeed capable of generating such distributions in cases where they genuinely exist; one clear example is that of the cubic phase of cristobalite \cite{Tucker:2001vg}.} Second, the fitted variance of these Gaussian distributions increases linearly with temperature, exactly as one would expect for a harmonic oscillator. We conclude that, to the extent to which anharmonicity is important in this material, it is \textit{completely} described within the renormalised phonon approximation taken to fourth order. In other words, although fourth-order interactions seen by individual phonon modes may limit their amplitude at high temperature, there is an insufficient number of such modes with near-zero harmonic terms to make an appreciable difference to the distribution of atomic positions. Thus overall the most important anharmonic interactions involve couplings between different phonons, as described by the renormalised phonon approximation taken to fourth order rather than the independent mode approximation. Furthermore, our results also rule out the significant of higher-order terms because they too would lead to a different temperature dependence. 

It is worth making a general comment here. It is very tempting to assume that because high temperatures lead to large-amplitude motions, they also lead to significant and unusual anharmonic effects. Actually this need not be so, given that large amplitudes are perfectly possible within the harmonic approximation. Our analysis here shows that high temperature does not necessarily produce unusual behaviour, such as envisaged in reference \onlinecite{Wendt:2019it} for example. Instead, at high temperature ScF$_3$ behaves as a typical harmonic crystal, or at least as one whose behaviour is only weakly perturbed by anharmonic effects. A similar conclusion was obtained from an RMC study of BiFeO$_3$ based on neutron total scattering data; in spite of very large atomic motions at high temperature, the average structure remains robustly constant across a wide range of temperatures and there are no unusual changes in atom distributions \cite{Du:2019ei}.

\section{Discussion and conclusions}\label{sec:discussion}
Our two key conclusions from the analysis of the RMC configurations discussed above are that that the NTE arises from a set of phonons with wave vectors around the lines of RUMs in reciprocal space and which are sufficiently extensive because of the relatively lower force constants associated with bending the bonds within the \ce{ScF6} coordination octahedra than in related systems, and that the transverse displacement of the F atoms, although associated with a quartic mode, can be well described by a Gaussian distribution whose width varies linearly with temperature, consistent with a renormalised harmonic phonon model. These results are closely related, since the RUMs are exactly the modes that will have small quadratic terms and in which the quartic terms are thus expected to be dominant. What they show is that the quasi-RUM vibrations make dominant contributions to the thermal expansion, and in particular to the NTE, of ScF$_3$. A similar conclusion was reached by the authors of reference \cite{Bird:2020dz} by a different type of analysis.%, demonstrating consistency with our results obtained by a pre-print version of this paper.

We need to state clearly that the deformations of the ScF$_6$ octahedra allowed within the quasi-RUMs do not in any way repudiate the importance of RUMs. It is a common misconception that the RUM model requires the octahedra to be very rigid (a mistake propagated in reference \onlinecite{Wendt:2019it}), but in fact quite the opposite is true, as has been discussed in detail recently in reference \onlinecite{Dove:2019dl}. The basic RUM model has always considered the structural polyhedra to have \textit{finite}, rather than infinite, rigidity, which in the original application of the RUM model to displacive phase transitions is directly associated with the phase transition temperature \cite{Dove:1992db, Dove:1995cw, Dove:1996tc, Gambhir:1997gv, Dove:1999uu}. In the same vein, the RUM model itself does not presuppose that any rigidity of the polyhedra arises only from covalent bonding. Polyhedral rigidity certainly can arise from Coulomb interactions or steric hindrance effects between anions within the structural polyhedra. In this sense our work here is also consistent with the viewpoint of reference \cite{Tkachenko:2020}, which discusses RUMs and quasi-RUMs in the context of a model with ionic forces.

The point is this: the fact that the RUMs are restricted to wave vectors lying on lines in reciprocal space means that they are a vanishingly small fraction of the total number of phonon modes and therefore alone they cannot give overall NTE. Instead, to get overall NTE it is necessary that there is a sufficiently-large number of low-frequency RUM-like phonon modes---quasi-RUMs---close to the wave vectors of the RUMs. This is possible if the polyhedra (in this case the ScF$_6$ octahedra) have some flexibility. Thus we propose firstly that the existence of the RUMs gives a set of phonons with the necessary low frequencies and appropriate mode eigenvectors for the tension effect to give NTE, and secondly that the flexibility of the polyhedra allows the contribution from the quasi-RUMs to spread across a sufficiently large volume in reciprocal space to have enough thermodynamic weighting to give an overall NTE. This explains concisely why there is NTE in ScF$_3$ but only positive thermal expansion in other perovskites such as SrTiO$_3$; both materials have stiff cation--anion (Sc--F or Ti--O) bonds, but the TiO$_6$ octahedra are more resistant to bond-bending distortion than the ScF$_6$ octahedra. This interpretation provides a plausible and reasonable explanation of the origin of NTE in ScF$_3$: one that is predictive, that is based on standard concepts in condensed matter physics, and that is consistent with previous experimental data and simulations.% \footnote{In this sense we cannot accept that it is valid to reject the value of phonon dispersion curves in favour of independent Einstein oscillators, as in the ``Entropic elasticity'' ideas of reference \onlinecite{Wendt:2019it}, which appear to be founded on errors in data, as discussed previously in footnote \onlinecite{Note3}.}.

%Although they will give some distortion of the ScF$_6$ octahedra, if the energy cost is not high the frequencies will not rise too fast on moving away from the RUM wave vectors. This is accomplished if the polyhedra (in this case the ScF$_6$ octahedra) are allowed some flexibility, which will keep the phonons near the RUM line at lower frequency and hence with more population significance within the thermodynamic functions. Thus we propose firstly that the existence of the RUMs gives a set of phonons with the necessary low frequencies and appropriate mode eigenvectors for the tension effect to give NTE, and secondly that the flexibility of the polyhedra allows the contribution from the RUMs to spread across a sufficiently large volume to give NTE. This explains concisely why we have NTE in ScF$_3$ but only positive thermal expansion in other perovskites such as SrTiO$_3$; both materials have stiff cation--anion (Sc--F or Ti--O) bonds, but the TiO$_6$ octahedra are more resistant to bond-bending distortion than the ScF$_6$ octahedra.

In conclusion, our real-space analysis of ScF$_3$, based on using the Reverse Monte Carlo method with neutron total scattering data to generate configurations of atoms over a wide range of temperatures, has allowed us to establish a quantitative view of the structural fluctuations associated with NTE. Comparison with a similar study of SrTiO$_3$, together with comparisons of published phonon dispersion curves, shows the importance of RUMs in giving rise to NTE, but that is also necessary to have some degree of polyhedral distortion to spread the contributions to NTE across a wider range of wave vectors than those associated with the pure RUMs. Fluorinated octahedra have bonds that are as stiff as in their oxygenated counterparts, but have more bond-bending flexibility. On this basis we suggest that in the search for materials with large negative coefficients of thermal expansion, fluorinated analogues of other oxides with NTE---one example being ZnF$_2$ as an analogue of the rutile phase of TiO$_2$ \cite{Wang:2015eo}---might prove to be particularly fertile \footnote{The research group of Angus Wilkinson has explored several fluorides with crystal structures analogous to ScF$_3$, including materials with various levels of doping \cite{Morelock:2013gi,Morelock:2014kw,Hancock:2015hl,Morelock:2015hh,Hester:2018ig}. Many examples display phase transitions involving rotations of octahedra and which appear to be continuous (second-order). These phase transitions will be accompanied by the softening of the RUM phonons on cooling towards the transition temperature, through the types of anharmonic interactions discussed in this paper as described by equation \ref{eq:rnpt}. We have shown that such a variation of the renormalised phonon frequency with temperature will lead to a reduction, or even elimination, of NTE \cite{Fang:2014cp}, as also demonstrated in recent calculations \cite{vanRoekeghem:2016kd,Oba:2019bi}.}.

%This is not to repudiate the importance of RUMs, but rather to emphasise---especially in perovskite-like systems such as \ce{ScF3}, where RUMs only occur at specific wavevectors---that contributions to the thermal expansion from other vibrational modes are likely to be important. Indeed, the opposite effect to this, in some sense, is known in some Prussian blues: in these perovskite-like structures RUMs are possible at any wavevector, but contributions from non-RUM modes ensure that the overall coefficient of thermal expansion remains positive.\cite{{Matsuda:2009hg}, {Adak:2011kr}}

%Our results emphasise the importance of real-space descriptions of negative thermal expansion materials. In particular, we anticipate that reverse Monte Carlo modelling will help to explain the behaviour of many materials in this remarkable family.

\section*{Data availability}

Original data sets are available directly from ISIS with Digital Object Identifier 10.5286/ISIS.E.RB1510519 \cite{data}. Corrected data and atomic configurations are available on request from the corresponding author.

\section*{Acknowledgements}

We are grateful to ISIS for provision of neutron beam time, supported under proposal number RB1510519. We also appreciate help from Helen Playford (ISIS) in preparation for the experimental beam time. J.D. is grateful to the China Scholarship Council and Queen Mary University of London for financial support. This research utilised the following computing resources: a) Queen Mary's Apocrita HPC facility (DOI: 10.5281/zenodo.438045), supported by QMUL Research-IT and funded by EPSRC grants EP/K000128/1 and EP/K000233/1 (M.T.D.); b) Midlands Plus Tier-2 HPC facility, funded by EPSRC grant EP/P020232/1 (M.T.D.).

%\section*{Author contributions}
%M.T.D. and A.E.P. designed the study. J.D. and M.T.D carried out the experiments, with support from D.A.K. and M.G.T.. J.D. performed the data reduction with support from D.A.K. and M.G.T.. J.D. performed the RMC analysis with support from M.T.D. and A.E.P.. MD simulations were performed by Z.W. and M.T.D.. The manuscript was written by M.T.D. and revised with input from A.E.P., D.A.K., and M.G.T..
%
%\section*{Competing interests}
%The authors declare no competing interests.
%
%\section*{Materials \& Correspondence}
%All correspondence and material requests should be addressed to M.T.D. by email, martin.dove@icloud.com.

%\bibliography{juan}

\begin{thebibliography}{87}%
\makeatletter
\providecommand \@ifxundefined [1]{%
 \@ifx{#1\undefined}
}%
\providecommand \@ifnum [1]{%
 \ifnum #1\expandafter \@firstoftwo
 \else \expandafter \@secondoftwo
 \fi
}%
\providecommand \@ifx [1]{%
 \ifx #1\expandafter \@firstoftwo
 \else \expandafter \@secondoftwo
 \fi
}%
\providecommand \natexlab [1]{#1}%
\providecommand \enquote  [1]{``#1''}%
\providecommand \bibnamefont  [1]{#1}%
\providecommand \bibfnamefont [1]{#1}%
\providecommand \citenamefont [1]{#1}%
\providecommand \href@noop [0]{\@secondoftwo}%
\providecommand \href [0]{\begingroup \@sanitize@url \@href}%
\providecommand \@href[1]{\@@startlink{#1}\@@href}%
\providecommand \@@href[1]{\endgroup#1\@@endlink}%
\providecommand \@sanitize@url [0]{\catcode `\\12\catcode `\$12\catcode
  `\&12\catcode `\#12\catcode `\^12\catcode `\_12\catcode `\%12\relax}%
\providecommand \@@startlink[1]{}%
\providecommand \@@endlink[0]{}%
\providecommand \url  [0]{\begingroup\@sanitize@url \@url }%
\providecommand \@url [1]{\endgroup\@href {#1}{\urlprefix }}%
\providecommand \urlprefix  [0]{URL }%
\providecommand \Eprint [0]{\href }%
\providecommand \doibase [0]{http://dx.doi.org/}%
\providecommand \selectlanguage [0]{\@gobble}%
\providecommand \bibinfo  [0]{\@secondoftwo}%
\providecommand \bibfield  [0]{\@secondoftwo}%
\providecommand \translation [1]{[#1]}%
\providecommand \BibitemOpen [0]{}%
\providecommand \bibitemStop [0]{}%
\providecommand \bibitemNoStop [0]{.\EOS\space}%
\providecommand \EOS [0]{\spacefactor3000\relax}%
\providecommand \BibitemShut  [1]{\csname bibitem#1\endcsname}%
\let\auto@bib@innerbib\@empty
%</preamble>
\bibitem [{\citenamefont {Dove}\ and\ \citenamefont
  {Fang}(2016)}]{Dove:2016bv}%
  \BibitemOpen
  \bibfield  {author} {\bibinfo {author} {\bibfnamefont {Martin~T}\
  \bibnamefont {Dove}}\ and\ \bibinfo {author} {\bibfnamefont {Hong}\
  \bibnamefont {Fang}},\ }\bibfield  {title} {\enquote {\bibinfo {title}
  {{Negative thermal expansion and associated anomalous physical properties:
  review of the lattice dynamics theoretical foundation}},}\ }\href {\doibase
  10.1088/0034-4885/79/6/066503} {\bibfield  {journal} {\bibinfo  {journal}
  {Reports on Progress in Physics}\ }\textbf {\bibinfo {volume} {79}},\
  \bibinfo {pages} {066503} (\bibinfo {year} {2016})}\BibitemShut {NoStop}%
\bibitem [{\citenamefont {Chen}\ \emph {et~al.}(2015)\citenamefont {Chen},
  \citenamefont {Hu}, \citenamefont {Deng},\ and\ \citenamefont
  {Xing}}]{Chen:2015hn}%
  \BibitemOpen
  \bibfield  {author} {\bibinfo {author} {\bibfnamefont {Jun}\ \bibnamefont
  {Chen}}, \bibinfo {author} {\bibfnamefont {Lei}\ \bibnamefont {Hu}}, \bibinfo
  {author} {\bibfnamefont {Jinxia}\ \bibnamefont {Deng}}, \ and\ \bibinfo
  {author} {\bibfnamefont {Xianran}\ \bibnamefont {Xing}},\ }\bibfield  {title}
  {\enquote {\bibinfo {title} {{Negative thermal expansion in functional
  materials: controllable thermal expansion by chemical modifications.}}}\
  }\href {\doibase 10.1039/C4CS00461B} {\bibfield  {journal} {\bibinfo
  {journal} {Chemical Society Reviews}\ }\textbf {\bibinfo {volume} {44}},\
  \bibinfo {pages} {3522--3567} (\bibinfo {year} {2015})}\BibitemShut {NoStop}%
\bibitem [{\citenamefont {Lind}(2012)}]{Lind:2012jwa}%
  \BibitemOpen
  \bibfield  {author} {\bibinfo {author} {\bibfnamefont {Cora}\ \bibnamefont
  {Lind}},\ }\bibfield  {title} {\enquote {\bibinfo {title} {{Two Decades of
  Negative Thermal Expansion Research: Where Do We Stand?}}}\ }\href {\doibase
  10.3390/ma5061125} {\bibfield  {journal} {\bibinfo  {journal} {Materials}\
  }\textbf {\bibinfo {volume} {5}},\ \bibinfo {pages} {1125--1154} (\bibinfo
  {year} {2012})}\BibitemShut {NoStop}%
\bibitem [{\citenamefont {Romao}\ \emph {et~al.}(2013)\citenamefont {Romao},
  \citenamefont {Miller}, \citenamefont {Whitman}, \citenamefont {White},\ and\
  \citenamefont {Marinkovic}}]{Romao:2013ch}%
  \BibitemOpen
  \bibfield  {author} {\bibinfo {author} {\bibfnamefont {C~P}\ \bibnamefont
  {Romao}}, \bibinfo {author} {\bibfnamefont {K~J}\ \bibnamefont {Miller}},
  \bibinfo {author} {\bibfnamefont {C~A}\ \bibnamefont {Whitman}}, \bibinfo
  {author} {\bibfnamefont {M~A}\ \bibnamefont {White}}, \ and\ \bibinfo
  {author} {\bibfnamefont {B~A}\ \bibnamefont {Marinkovic}},\ }\bibfield
  {title} {\enquote {\bibinfo {title} {{Negative thermal expansion
  (thermomiotic) materials.}}}\ }in\ \href {\doibase
  10.1016/B978-0-08-097774-4.00425-3} {\emph {\bibinfo {booktitle}
  {Comprehensive Inorganic Chemistry II: From elements to applications}}}\
  (\bibinfo  {publisher} {Elsevier},\ \bibinfo {year} {2013})\ pp.\ \bibinfo
  {pages} {127--151}\BibitemShut {NoStop}%
\bibitem [{\citenamefont {Takenaka}(2012)}]{Takenaka:2012cv}%
  \BibitemOpen
  \bibfield  {author} {\bibinfo {author} {\bibfnamefont {Koshi}\ \bibnamefont
  {Takenaka}},\ }\bibfield  {title} {\enquote {\bibinfo {title} {{Negative
  thermal expansion materials: technological key for control of thermal
  expansion.}}}\ }\href {\doibase 10.1088/1468-6996/13/1/013001} {\bibfield
  {journal} {\bibinfo  {journal} {Science and Technology of Advanced
  Materials}\ }\textbf {\bibinfo {volume} {13}},\ \bibinfo {pages} {013001}
  (\bibinfo {year} {2012})}\BibitemShut {NoStop}%
\bibitem [{\citenamefont {Barrera}\ \emph {et~al.}(2005)\citenamefont
  {Barrera}, \citenamefont {Bruno}, \citenamefont {Barron},\ and\ \citenamefont
  {Allan}}]{Barrera:2005dt}%
  \BibitemOpen
  \bibfield  {author} {\bibinfo {author} {\bibfnamefont {G~D}\ \bibnamefont
  {Barrera}}, \bibinfo {author} {\bibfnamefont {JAO}\ \bibnamefont {Bruno}},
  \bibinfo {author} {\bibfnamefont {THK}\ \bibnamefont {Barron}}, \ and\
  \bibinfo {author} {\bibfnamefont {N~L}\ \bibnamefont {Allan}},\ }\bibfield
  {title} {\enquote {\bibinfo {title} {Negative thermal expansion},}\ }\href
  {\doibase 10.1088/0953-8984/17/4/R03} {\bibfield  {journal} {\bibinfo
  {journal} {Journal of Physics: Condensed Matter}\ }\textbf {\bibinfo {volume}
  {17}},\ \bibinfo {pages} {R217--R252} (\bibinfo {year} {2005})}\BibitemShut
  {NoStop}%
\bibitem [{\citenamefont {Liang}(2010)}]{Liang:2010uj}%
  \BibitemOpen
  \bibfield  {author} {\bibinfo {author} {\bibfnamefont {Er-Jun}\ \bibnamefont
  {Liang}},\ }\bibfield  {title} {\enquote {\bibinfo {title} {{Negative thermal
  expansion materials and their applications: A survey of recent patents.}}}\
  }\href {http://eurekaselect.com/93812} {\bibfield  {journal} {\bibinfo
  {journal} {Recent Patents on Materials Science}\ }\textbf {\bibinfo {volume}
  {3}},\ \bibinfo {pages} {106--128} (\bibinfo {year} {2010})}\BibitemShut
  {NoStop}%
\bibitem [{\citenamefont {Ren}\ \emph {et~al.}(2018)\citenamefont {Ren},
  \citenamefont {Das}, \citenamefont {Tran}, \citenamefont {Ngo},\ and\
  \citenamefont {Xie}}]{Ren:2018ie}%
  \BibitemOpen
  \bibfield  {author} {\bibinfo {author} {\bibfnamefont {Xin}\ \bibnamefont
  {Ren}}, \bibinfo {author} {\bibfnamefont {Raj}\ \bibnamefont {Das}}, \bibinfo
  {author} {\bibfnamefont {Phuong}\ \bibnamefont {Tran}}, \bibinfo {author}
  {\bibfnamefont {Tuan~Duc}\ \bibnamefont {Ngo}}, \ and\ \bibinfo {author}
  {\bibfnamefont {Yi~Min}\ \bibnamefont {Xie}},\ }\bibfield  {title} {\enquote
  {\bibinfo {title} {{Auxetic metamaterials and structures: a review}},}\
  }\href {\doibase 10.1088/1361-665X/aaa61c} {\bibfield  {journal} {\bibinfo
  {journal} {Smart Materials and Structures}\ }\textbf {\bibinfo {volume}
  {27}},\ \bibinfo {pages} {023001--39} (\bibinfo {year} {2018})}\BibitemShut
  {NoStop}%
\bibitem [{\citenamefont {Fang}\ and\ \citenamefont
  {Dove}(2013)}]{Fang:2013gp}%
  \BibitemOpen
  \bibfield  {author} {\bibinfo {author} {\bibfnamefont {Hong}\ \bibnamefont
  {Fang}}\ and\ \bibinfo {author} {\bibfnamefont {Martin~T}\ \bibnamefont
  {Dove}},\ }\bibfield  {title} {\enquote {\bibinfo {title} {{Pressure-induced
  softening as a common feature of framework structures with negative thermal
  expansion.}}}\ }\href {\doibase 10.1103/PhysRevB.87.214109} {\bibfield
  {journal} {\bibinfo  {journal} {Physical Review B}\ }\textbf {\bibinfo
  {volume} {87}},\ \bibinfo {pages} {214109} (\bibinfo {year}
  {2013})}\BibitemShut {NoStop}%
\bibitem [{\citenamefont {Fang}\ \emph
  {et~al.}(2013{\natexlab{a}})\citenamefont {Fang}, \citenamefont {Phillips},
  \citenamefont {Dove}, \citenamefont {Tucker},\ and\ \citenamefont
  {Goodwin}}]{Fang:2013fj}%
  \BibitemOpen
  \bibfield  {author} {\bibinfo {author} {\bibfnamefont {Hong}\ \bibnamefont
  {Fang}}, \bibinfo {author} {\bibfnamefont {Anthony~E}\ \bibnamefont
  {Phillips}}, \bibinfo {author} {\bibfnamefont {Martin~T}\ \bibnamefont
  {Dove}}, \bibinfo {author} {\bibfnamefont {Matthew~G}\ \bibnamefont
  {Tucker}}, \ and\ \bibinfo {author} {\bibfnamefont {Andrew~L}\ \bibnamefont
  {Goodwin}},\ }\bibfield  {title} {\enquote {\bibinfo {title}
  {{Temperature-dependent pressure-induced softening in Zn(CN)$_2$.}}}\ }\href
  {\doibase 10.1103/PhysRevB.88.144103} {\bibfield  {journal} {\bibinfo
  {journal} {Physical Review B}\ }\textbf {\bibinfo {volume} {88}},\ \bibinfo
  {pages} {144103} (\bibinfo {year} {2013}{\natexlab{a}})}\BibitemShut
  {NoStop}%
\bibitem [{\citenamefont {Wei}\ \emph {et~al.}(2020)\citenamefont {Wei},
  \citenamefont {Tan}, \citenamefont {Cai}, \citenamefont {Phillips},
  \citenamefont {{da Silva}}, \citenamefont {Kibble},\ and\ \citenamefont
  {Dove}}]{Wei:2020}%
  \BibitemOpen
  \bibfield  {author} {\bibinfo {author} {\bibfnamefont {Zhongsheng}\
  \bibnamefont {Wei}}, \bibinfo {author} {\bibfnamefont {Lei}\ \bibnamefont
  {Tan}}, \bibinfo {author} {\bibfnamefont {Guanqun}\ \bibnamefont {Cai}},
  \bibinfo {author} {\bibfnamefont {Anthony~E}\ \bibnamefont {Phillips}},
  \bibinfo {author} {\bibfnamefont {Ivan}\ \bibnamefont {{da Silva}}}, \bibinfo
  {author} {\bibfnamefont {Mark~G}\ \bibnamefont {Kibble}}, \ and\ \bibinfo
  {author} {\bibfnamefont {Martin~T}\ \bibnamefont {Dove}},\ }\bibfield
  {title} {\enquote {\bibinfo {title} {Colossal pressure-induced softening in
  scandium fluoride},}\ }\href {\doibase 10.1103/PhysRevLett.124.255502}
  {\bibfield  {journal} {\bibinfo  {journal} {Physical Review Letters}\
  }\textbf {\bibinfo {volume} {124}},\ \bibinfo {pages} {255502} (\bibinfo
  {year} {2020})}\BibitemShut {NoStop}%
\bibitem [{\citenamefont {Mittal}\ \emph {et~al.}(2018)\citenamefont {Mittal},
  \citenamefont {Gupta},\ and\ \citenamefont {Chaplot}}]{Mittal:2018fy}%
  \BibitemOpen
  \bibfield  {author} {\bibinfo {author} {\bibfnamefont {R}~\bibnamefont
  {Mittal}}, \bibinfo {author} {\bibfnamefont {M~K}\ \bibnamefont {Gupta}}, \
  and\ \bibinfo {author} {\bibfnamefont {S~L}\ \bibnamefont {Chaplot}},\
  }\bibfield  {title} {\enquote {\bibinfo {title} {{Phonons and anomalous
  thermal expansion behaviour in crystalline solids}},}\ }\href {\doibase
  10.1016/j.pmatsci.2017.10.002} {\bibfield  {journal} {\bibinfo  {journal}
  {Progress in Materials Science}\ }\textbf {\bibinfo {volume} {92}},\ \bibinfo
  {pages} {360--445} (\bibinfo {year} {2018})}\BibitemShut {NoStop}%
\bibitem [{Note1()}]{Note1}%
  \BibitemOpen
  \bibinfo {note} {It may be argued that since, in comparison with the
  perovskite structure, ScF$_3$ has no A-site cation, it should have more
  flexibility for rotational motions and hence for the tension effect to
  operate. However, this is a real-space intuition that doesn't necessarily
  correspond directly with what is really found. Since vibrations are correctly
  resolved into a summation of normal modes with wave vectors in reciprocal
  space, any effects of the A-site cation should be interpreted in terms of
  their effect on the frequencies of the relevant phonon modes. And here we see
  that the effect is not to block the motion at all. First, comparing for
  ScF$_3$ \cite {Li:2011dn, Handunkanda:2015dc, Oba:2019bi, Handunkanda:2015dc}
  and SrTiO$_3$ \cite {Stirling:1972ul} the values of the lowest-frequency
  modes along the $M$--$R$ directions in reciprocal space, as defined later, we
  find very similar frequency values and hence the capacity for similar RUM
  amplitudes. Furthermore, in many cubic perovskites there is a softening of
  the RUM phonon frequencies on cooling towards a displacive phase transition,
  which will increase the RUM amplitude.}\BibitemShut {Stop}%
\bibitem [{\citenamefont {Greve}\ \emph {et~al.}(2010)\citenamefont {Greve},
  \citenamefont {Martin}, \citenamefont {Lee}, \citenamefont {Chupas},
  \citenamefont {Chapman},\ and\ \citenamefont {Wilkinson}}]{Greve:2010bu}%
  \BibitemOpen
  \bibfield  {author} {\bibinfo {author} {\bibfnamefont {Benjamin~K}\
  \bibnamefont {Greve}}, \bibinfo {author} {\bibfnamefont {Kenneth~L}\
  \bibnamefont {Martin}}, \bibinfo {author} {\bibfnamefont {Peter~L}\
  \bibnamefont {Lee}}, \bibinfo {author} {\bibfnamefont {Peter~J}\ \bibnamefont
  {Chupas}}, \bibinfo {author} {\bibfnamefont {Karena~W}\ \bibnamefont
  {Chapman}}, \ and\ \bibinfo {author} {\bibfnamefont {Angus~P}\ \bibnamefont
  {Wilkinson}},\ }\bibfield  {title} {\enquote {\bibinfo {title} {{Pronounced
  negative thermal expansion from a simple structure: Cubic ScF$_3$.}}}\ }\href
  {\doibase 10.1021/ja106711v} {\bibfield  {journal} {\bibinfo  {journal}
  {Journal of the American Chemical Society}\ }\textbf {\bibinfo {volume}
  {132}},\ \bibinfo {pages} {15496--15498} (\bibinfo {year}
  {2010})}\BibitemShut {NoStop}%
\bibitem [{\citenamefont {Hu}\ \emph {et~al.}(2016)\citenamefont {Hu},
  \citenamefont {Chen}, \citenamefont {Sanson}, \citenamefont {Wu},
  \citenamefont {Guglieri~Rodriguez}, \citenamefont {Olivi}, \citenamefont
  {Ren}, \citenamefont {Fan}, \citenamefont {Deng},\ and\ \citenamefont
  {Xing}}]{Hu:2016it}%
  \BibitemOpen
  \bibfield  {author} {\bibinfo {author} {\bibfnamefont {Lei}\ \bibnamefont
  {Hu}}, \bibinfo {author} {\bibfnamefont {Jun}\ \bibnamefont {Chen}}, \bibinfo
  {author} {\bibfnamefont {Andrea}\ \bibnamefont {Sanson}}, \bibinfo {author}
  {\bibfnamefont {Hui}\ \bibnamefont {Wu}}, \bibinfo {author} {\bibfnamefont
  {Clara}\ \bibnamefont {Guglieri~Rodriguez}}, \bibinfo {author} {\bibfnamefont
  {Luca}\ \bibnamefont {Olivi}}, \bibinfo {author} {\bibfnamefont {Yang}\
  \bibnamefont {Ren}}, \bibinfo {author} {\bibfnamefont {Longlong}\
  \bibnamefont {Fan}}, \bibinfo {author} {\bibfnamefont {Jinxia}\ \bibnamefont
  {Deng}}, \ and\ \bibinfo {author} {\bibfnamefont {Xianran}\ \bibnamefont
  {Xing}},\ }\bibfield  {title} {\enquote {\bibinfo {title} {{New Insights into
  the Negative Thermal Expansion: Direct Experimental Evidence for the
  ``Guitar-String'' Effect in Cubic ScF$_3$.}}}\ }\href {\doibase
  10.1021/jacs.6b02370} {\bibfield  {journal} {\bibinfo  {journal} {Journal of
  the American Chemical Society}\ }\textbf {\bibinfo {volume} {138}},\ \bibinfo
  {pages} {8320--8323} (\bibinfo {year} {2016})}\BibitemShut {NoStop}%
\bibitem [{\citenamefont {Li}\ \emph {et~al.}(2011)\citenamefont {Li},
  \citenamefont {Tang}, \citenamefont {Mu{\~n}oz}, \citenamefont {Keith},
  \citenamefont {Tracy}, \citenamefont {Abernathy},\ and\ \citenamefont
  {Fultz}}]{Li:2011dn}%
  \BibitemOpen
  \bibfield  {author} {\bibinfo {author} {\bibfnamefont {Chen~W}\ \bibnamefont
  {Li}}, \bibinfo {author} {\bibfnamefont {Xiaoli}\ \bibnamefont {Tang}},
  \bibinfo {author} {\bibfnamefont {J~A}\ \bibnamefont {Mu{\~n}oz}}, \bibinfo
  {author} {\bibfnamefont {J~B}\ \bibnamefont {Keith}}, \bibinfo {author}
  {\bibfnamefont {S~J}\ \bibnamefont {Tracy}}, \bibinfo {author} {\bibfnamefont
  {D~L}\ \bibnamefont {Abernathy}}, \ and\ \bibinfo {author} {\bibfnamefont
  {B}~\bibnamefont {Fultz}},\ }\bibfield  {title} {\enquote {\bibinfo {title}
  {{Structural relationship between negative thermal expansion and quartic
  anharmonicity of cubic ScF$_3$.}}}\ }\href {\doibase
  10.1103/PhysRevLett.107.195504} {\bibfield  {journal} {\bibinfo  {journal}
  {Physical Review Letters}\ }\textbf {\bibinfo {volume} {107}},\ \bibinfo
  {pages} {195504} (\bibinfo {year} {2011})}\BibitemShut {NoStop}%
\bibitem [{\citenamefont {Hibble}\ \emph
  {et~al.}(2002{\natexlab{a}})\citenamefont {Hibble}, \citenamefont {Cheyne},
  \citenamefont {Hannon},\ and\ \citenamefont {Eversfield}}]{Hibble:2002dd}%
  \BibitemOpen
  \bibfield  {author} {\bibinfo {author} {\bibfnamefont {Simon~J}\ \bibnamefont
  {Hibble}}, \bibinfo {author} {\bibfnamefont {Simon~M}\ \bibnamefont
  {Cheyne}}, \bibinfo {author} {\bibfnamefont {Alex~C}\ \bibnamefont {Hannon}},
  \ and\ \bibinfo {author} {\bibfnamefont {Sharon~G}\ \bibnamefont
  {Eversfield}},\ }\bibfield  {title} {\enquote {\bibinfo {title} {{Beyond
  Bragg scattering:~ the structure of AgCN determined from total neutron
  diffraction.}}}\ }\href@noop {} {\bibfield  {journal} {\bibinfo  {journal}
  {Inorganic Chemistry}\ }\textbf {\bibinfo {volume} {41}},\ \bibinfo {pages}
  {1042--1044} (\bibinfo {year} {2002}{\natexlab{a}})}\BibitemShut {NoStop}%
\bibitem [{\citenamefont {Hibble}\ \emph
  {et~al.}(2002{\natexlab{b}})\citenamefont {Hibble}, \citenamefont {Cheyne},
  \citenamefont {Hannon},\ and\ \citenamefont {Eversfield}}]{Hibble:2002fv}%
  \BibitemOpen
  \bibfield  {author} {\bibinfo {author} {\bibfnamefont {Simon~J}\ \bibnamefont
  {Hibble}}, \bibinfo {author} {\bibfnamefont {Simon~M}\ \bibnamefont
  {Cheyne}}, \bibinfo {author} {\bibfnamefont {Alex~C}\ \bibnamefont {Hannon}},
  \ and\ \bibinfo {author} {\bibfnamefont {Sharon~G}\ \bibnamefont
  {Eversfield}},\ }\bibfield  {title} {\enquote {\bibinfo {title} {{CuCN:~ a
  polymorphic material. Structure of one form determined from total neutron
  diffraction.}}}\ }\href@noop {} {\bibfield  {journal} {\bibinfo  {journal}
  {Inorganic Chemistry}\ }\textbf {\bibinfo {volume} {41}},\ \bibinfo {pages}
  {4990--4992} (\bibinfo {year} {2002}{\natexlab{b}})}\BibitemShut {NoStop}%
\bibitem [{\citenamefont {Tucker}\ \emph {et~al.}(2005)\citenamefont {Tucker},
  \citenamefont {Goodwin}, \citenamefont {Dove}, \citenamefont {Keen},
  \citenamefont {Wells},\ and\ \citenamefont {Evans}}]{Tucker:2005hk}%
  \BibitemOpen
  \bibfield  {author} {\bibinfo {author} {\bibfnamefont {Matthew~G}\
  \bibnamefont {Tucker}}, \bibinfo {author} {\bibfnamefont {Andrew~L}\
  \bibnamefont {Goodwin}}, \bibinfo {author} {\bibfnamefont {Martin~T}\
  \bibnamefont {Dove}}, \bibinfo {author} {\bibfnamefont {David~A}\
  \bibnamefont {Keen}}, \bibinfo {author} {\bibfnamefont {Stephen~A}\
  \bibnamefont {Wells}}, \ and\ \bibinfo {author} {\bibfnamefont {John S~O}\
  \bibnamefont {Evans}},\ }\bibfield  {title} {\enquote {\bibinfo {title}
  {{Negative thermal expansion in ZrW$_2$O$_8$: mechanisms, rigid unit modes,
  and neutron total scattering.}}}\ }\href@noop {} {\bibfield  {journal}
  {\bibinfo  {journal} {Physical Review Letters}\ }\textbf {\bibinfo {volume}
  {95}},\ \bibinfo {pages} {255501} (\bibinfo {year} {2005})}\BibitemShut
  {NoStop}%
\bibitem [{\citenamefont {Chapman}\ \emph {et~al.}(2005)\citenamefont
  {Chapman}, \citenamefont {Chupas},\ and\ \citenamefont
  {Kepert}}]{Chapman:2005fj}%
  \BibitemOpen
  \bibfield  {author} {\bibinfo {author} {\bibfnamefont {Karena~W}\
  \bibnamefont {Chapman}}, \bibinfo {author} {\bibfnamefont {Peter~J}\
  \bibnamefont {Chupas}}, \ and\ \bibinfo {author} {\bibfnamefont {Cameron~J}\
  \bibnamefont {Kepert}},\ }\bibfield  {title} {\enquote {\bibinfo {title}
  {{Direct observation of a transverse vibrational mechanism for negative
  thermal expansion in Zn(CN)$_2$:~ an atomic pair distribution function
  analysis.}}}\ }\href {\doibase 10.1021/ja055197f} {\bibfield  {journal}
  {\bibinfo  {journal} {Journal of the American Chemical Society}\ }\textbf
  {\bibinfo {volume} {127}},\ \bibinfo {pages} {15630--15636} (\bibinfo {year}
  {2005})}\BibitemShut {NoStop}%
\bibitem [{\citenamefont {Tucker}\ \emph
  {et~al.}(2007{\natexlab{a}})\citenamefont {Tucker}, \citenamefont {Keen},
  \citenamefont {Evans},\ and\ \citenamefont {Dove}}]{Tucker:2007gk}%
  \BibitemOpen
  \bibfield  {author} {\bibinfo {author} {\bibfnamefont {Matthew~G}\
  \bibnamefont {Tucker}}, \bibinfo {author} {\bibfnamefont {David~A}\
  \bibnamefont {Keen}}, \bibinfo {author} {\bibfnamefont {John S~O}\
  \bibnamefont {Evans}}, \ and\ \bibinfo {author} {\bibfnamefont {Martin~T}\
  \bibnamefont {Dove}},\ }\bibfield  {title} {\enquote {\bibinfo {title}
  {{Local structure in ZrW$_2$O$_8$ from neutron total scattering.}}}\
  }\href@noop {} {\bibfield  {journal} {\bibinfo  {journal} {Journal of
  Physics: Condensed Matter}\ }\textbf {\bibinfo {volume} {19}},\ \bibinfo
  {pages} {335215} (\bibinfo {year} {2007}{\natexlab{a}})}\BibitemShut
  {NoStop}%
\bibitem [{\citenamefont {Dapiaggi}\ \emph {et~al.}(2008)\citenamefont
  {Dapiaggi}, \citenamefont {Kim}, \citenamefont {Bo{\v z}in}, \citenamefont
  {Billinge},\ and\ \citenamefont {Artioli}}]{Dapiaggi:2008et}%
  \BibitemOpen
  \bibfield  {author} {\bibinfo {author} {\bibfnamefont {Monica}\ \bibnamefont
  {Dapiaggi}}, \bibinfo {author} {\bibfnamefont {HyunJeong}\ \bibnamefont
  {Kim}}, \bibinfo {author} {\bibfnamefont {Emil~S}\ \bibnamefont {Bo{\v
  z}in}}, \bibinfo {author} {\bibfnamefont {Simon J~L}\ \bibnamefont
  {Billinge}}, \ and\ \bibinfo {author} {\bibfnamefont {Gilberto}\ \bibnamefont
  {Artioli}},\ }\bibfield  {title} {\enquote {\bibinfo {title} {{Study of the
  negative thermal expansion of cuprite-type structures by means of
  temperature-dependent pair distribution function analysis: Preliminary
  results.}}}\ }\href@noop {} {\bibfield  {journal} {\bibinfo  {journal}
  {Journal of Physics and Chemistry of Solids}\ }\textbf {\bibinfo {volume}
  {69}},\ \bibinfo {pages} {2182--2186} (\bibinfo {year} {2008})}\BibitemShut
  {NoStop}%
\bibitem [{\citenamefont {Chapman}\ and\ \citenamefont
  {Chupas}(2009)}]{Chapman:2009fs}%
  \BibitemOpen
  \bibfield  {author} {\bibinfo {author} {\bibfnamefont {Karena~W}\
  \bibnamefont {Chapman}}\ and\ \bibinfo {author} {\bibfnamefont {Peter~J}\
  \bibnamefont {Chupas}},\ }\bibfield  {title} {\enquote {\bibinfo {title}
  {{Anomalous thermal expansion of cuprites: A combined high resolution pair
  distribution function and geometric analysis.}}}\ }\href@noop {} {\bibfield
  {journal} {\bibinfo  {journal} {Chemistry of Materials}\ }\textbf {\bibinfo
  {volume} {21}},\ \bibinfo {pages} {425--431} (\bibinfo {year}
  {2009})}\BibitemShut {NoStop}%
\bibitem [{\citenamefont {Hibble}\ \emph {et~al.}(2013)\citenamefont {Hibble},
  \citenamefont {Chippindale}, \citenamefont {Marelli}, \citenamefont
  {Kroeker}, \citenamefont {Michaelis}, \citenamefont {Greer}, \citenamefont
  {Aguiar}, \citenamefont {Bilb{\'e}}, \citenamefont {Barney},\ and\
  \citenamefont {Hannon}}]{Hibble:2013io}%
  \BibitemOpen
  \bibfield  {author} {\bibinfo {author} {\bibfnamefont {Simon~J}\ \bibnamefont
  {Hibble}}, \bibinfo {author} {\bibfnamefont {Ann~M}\ \bibnamefont
  {Chippindale}}, \bibinfo {author} {\bibfnamefont {Elena}\ \bibnamefont
  {Marelli}}, \bibinfo {author} {\bibfnamefont {Scott}\ \bibnamefont
  {Kroeker}}, \bibinfo {author} {\bibfnamefont {Vladimir~K}\ \bibnamefont
  {Michaelis}}, \bibinfo {author} {\bibfnamefont {Brandon~J}\ \bibnamefont
  {Greer}}, \bibinfo {author} {\bibfnamefont {Pedro~M}\ \bibnamefont {Aguiar}},
  \bibinfo {author} {\bibfnamefont {Edward~J}\ \bibnamefont {Bilb{\'e}}},
  \bibinfo {author} {\bibfnamefont {Emma~R}\ \bibnamefont {Barney}}, \ and\
  \bibinfo {author} {\bibfnamefont {Alex~C}\ \bibnamefont {Hannon}},\
  }\bibfield  {title} {\enquote {\bibinfo {title} {{Local and average structure
  in zinc cyanide: toward an understanding of the atomistic origin of negative
  thermal expansion.}}}\ }\href@noop {} {\bibfield  {journal} {\bibinfo
  {journal} {Journal of the American Chemical Society}\ }\textbf {\bibinfo
  {volume} {135}},\ \bibinfo {pages} {16478--16489} (\bibinfo {year}
  {2013})}\BibitemShut {NoStop}%
\bibitem [{\citenamefont {Bridges}\ \emph {et~al.}(2014)\citenamefont
  {Bridges}, \citenamefont {Keiber}, \citenamefont {Juhas}, \citenamefont
  {Billinge}, \citenamefont {Sutton}, \citenamefont {Wilde},\ and\
  \citenamefont {Kowach}}]{Bridges:2014bm}%
  \BibitemOpen
  \bibfield  {author} {\bibinfo {author} {\bibfnamefont {F}~\bibnamefont
  {Bridges}}, \bibinfo {author} {\bibfnamefont {T}~\bibnamefont {Keiber}},
  \bibinfo {author} {\bibfnamefont {P}~\bibnamefont {Juhas}}, \bibinfo {author}
  {\bibfnamefont {S~J~L}\ \bibnamefont {Billinge}}, \bibinfo {author}
  {\bibfnamefont {L}~\bibnamefont {Sutton}}, \bibinfo {author} {\bibfnamefont
  {J}~\bibnamefont {Wilde}}, \ and\ \bibinfo {author} {\bibfnamefont {Glen~R}\
  \bibnamefont {Kowach}},\ }\bibfield  {title} {\enquote {\bibinfo {title}
  {{Local vibrations and negative thermal expansion in ZrW$_2$O$_8$.}}}\
  }\href@noop {} {\bibfield  {journal} {\bibinfo  {journal} {Physical Review
  Letters}\ }\textbf {\bibinfo {volume} {112}},\ \bibinfo {pages} {045505}
  (\bibinfo {year} {2014})}\BibitemShut {NoStop}%
\bibitem [{\citenamefont {Wendt}\ \emph {et~al.}(2019)\citenamefont {Wendt},
  \citenamefont {Bozin}, \citenamefont {Neuefeind}, \citenamefont {Page},
  \citenamefont {Ku}, \citenamefont {Wang}, \citenamefont {Fultz},
  \citenamefont {Tkachenko},\ and\ \citenamefont {Zaliznyak}}]{Wendt:2019it}%
  \BibitemOpen
  \bibfield  {author} {\bibinfo {author} {\bibfnamefont {David}\ \bibnamefont
  {Wendt}}, \bibinfo {author} {\bibfnamefont {Emil}\ \bibnamefont {Bozin}},
  \bibinfo {author} {\bibfnamefont {Joerg}\ \bibnamefont {Neuefeind}}, \bibinfo
  {author} {\bibfnamefont {Katharine}\ \bibnamefont {Page}}, \bibinfo {author}
  {\bibfnamefont {Wei}\ \bibnamefont {Ku}}, \bibinfo {author} {\bibfnamefont
  {Limin}\ \bibnamefont {Wang}}, \bibinfo {author} {\bibfnamefont {Brent}\
  \bibnamefont {Fultz}}, \bibinfo {author} {\bibfnamefont {Alexei~V}\
  \bibnamefont {Tkachenko}}, \ and\ \bibinfo {author} {\bibfnamefont {Igor~A}\
  \bibnamefont {Zaliznyak}},\ }\bibfield  {title} {\enquote {\bibinfo {title}
  {{Entropic elasticity and negative thermal expansion in a simple cubic
  crystal}},}\ }\href@noop {} {\bibfield  {journal} {\bibinfo  {journal}
  {Science Advances}\ }\textbf {\bibinfo {volume} {5}},\ \bibinfo {pages}
  {eaay2748--8} (\bibinfo {year} {2019})}\BibitemShut {NoStop}%
\bibitem [{\citenamefont {Yang}\ \emph {et~al.}(2016)\citenamefont {Yang},
  \citenamefont {Tong}, \citenamefont {Lin}, \citenamefont {Guo}, \citenamefont
  {Zhang}, \citenamefont {Wang}, \citenamefont {Wu}, \citenamefont {Lin},
  \citenamefont {Huang}, \citenamefont {Xu}, \citenamefont {Song},\ and\
  \citenamefont {Sun}}]{Yang:2016ch}%
  \BibitemOpen
  \bibfield  {author} {\bibinfo {author} {\bibfnamefont {C}~\bibnamefont
  {Yang}}, \bibinfo {author} {\bibfnamefont {P}~\bibnamefont {Tong}}, \bibinfo
  {author} {\bibfnamefont {J~C}\ \bibnamefont {Lin}}, \bibinfo {author}
  {\bibfnamefont {X~G}\ \bibnamefont {Guo}}, \bibinfo {author} {\bibfnamefont
  {K}~\bibnamefont {Zhang}}, \bibinfo {author} {\bibfnamefont {M}~\bibnamefont
  {Wang}}, \bibinfo {author} {\bibfnamefont {Y}~\bibnamefont {Wu}}, \bibinfo
  {author} {\bibfnamefont {S}~\bibnamefont {Lin}}, \bibinfo {author}
  {\bibfnamefont {P~C}\ \bibnamefont {Huang}}, \bibinfo {author} {\bibfnamefont
  {W}~\bibnamefont {Xu}}, \bibinfo {author} {\bibfnamefont {W~H}\ \bibnamefont
  {Song}}, \ and\ \bibinfo {author} {\bibfnamefont {Y~P}\ \bibnamefont {Sun}},\
  }\bibfield  {title} {\enquote {\bibinfo {title} {{Size effects on negative
  thermal expansion in cubic ScF$_3$}},}\ }\href {\doibase 10.1063/1.4959083}
  {\bibfield  {journal} {\bibinfo  {journal} {Applied Physics Letters}\
  }\textbf {\bibinfo {volume} {109}},\ \bibinfo {pages} {023110} (\bibinfo
  {year} {2016})}\BibitemShut {NoStop}%
\bibitem [{\citenamefont {Hu}\ \emph {et~al.}(2018)\citenamefont {Hu},
  \citenamefont {Qin}, \citenamefont {Sanson}, \citenamefont {Huang},
  \citenamefont {Pan}, \citenamefont {Li}, \citenamefont {Sun}, \citenamefont
  {Wang}, \citenamefont {Guo}, \citenamefont {Aydemir}, \citenamefont {Ren},
  \citenamefont {Sun}, \citenamefont {Deng}, \citenamefont {Aquianti},
  \citenamefont {Rondinelli}, \citenamefont {Chen},\ and\ \citenamefont
  {Xing}}]{Hu:2018eq}%
  \BibitemOpen
  \bibfield  {author} {\bibinfo {author} {\bibfnamefont {Lei}\ \bibnamefont
  {Hu}}, \bibinfo {author} {\bibfnamefont {Feiyu}\ \bibnamefont {Qin}},
  \bibinfo {author} {\bibfnamefont {Andrea}\ \bibnamefont {Sanson}}, \bibinfo
  {author} {\bibfnamefont {Liang-Feng}\ \bibnamefont {Huang}}, \bibinfo
  {author} {\bibfnamefont {Zhaoo}\ \bibnamefont {Pan}}, \bibinfo {author}
  {\bibfnamefont {Quang}\ \bibnamefont {Li}}, \bibinfo {author} {\bibfnamefont
  {Qiang}\ \bibnamefont {Sun}}, \bibinfo {author} {\bibfnamefont
  {Lu}~\bibnamefont {Wang}}, \bibinfo {author} {\bibfnamefont {Fangmin}\
  \bibnamefont {Guo}}, \bibinfo {author} {\bibfnamefont {Umut}\ \bibnamefont
  {Aydemir}}, \bibinfo {author} {\bibfnamefont {Yang}\ \bibnamefont {Ren}},
  \bibinfo {author} {\bibfnamefont {Chengjun}\ \bibnamefont {Sun}}, \bibinfo
  {author} {\bibfnamefont {Jinxia}\ \bibnamefont {Deng}}, \bibinfo {author}
  {\bibfnamefont {Guiliana}\ \bibnamefont {Aquianti}}, \bibinfo {author}
  {\bibfnamefont {James~M}\ \bibnamefont {Rondinelli}}, \bibinfo {author}
  {\bibfnamefont {Jun}\ \bibnamefont {Chen}}, \ and\ \bibinfo {author}
  {\bibfnamefont {Xianran}\ \bibnamefont {Xing}},\ }\bibfield  {title}
  {\enquote {\bibinfo {title} {{Localized symmetry breaking for tuning thermal
  expansion in ScF$_3$ nanoscale frameworks}},}\ }\href {\doibase
  10.1021/jacs.8b00885} {\bibfield  {journal} {\bibinfo  {journal} {Journal of
  the Americna Chemical Society}\ }\textbf {\bibinfo {volume} {140}},\ \bibinfo
  {pages} {4477--4480} (\bibinfo {year} {2018})}\BibitemShut {NoStop}%
\bibitem [{\citenamefont {Bird}\ \emph {et~al.}(2020)\citenamefont {Bird},
  \citenamefont {Woodland-Scott}, \citenamefont {Hu}, \citenamefont {Wharmby},
  \citenamefont {Chen}, \citenamefont {Goodwin},\ and\ \citenamefont
  {Senn}}]{Bird:2020dz}%
  \BibitemOpen
  \bibfield  {author} {\bibinfo {author} {\bibfnamefont {T~A}\ \bibnamefont
  {Bird}}, \bibinfo {author} {\bibfnamefont {J}~\bibnamefont {Woodland-Scott}},
  \bibinfo {author} {\bibfnamefont {L}~\bibnamefont {Hu}}, \bibinfo {author}
  {\bibfnamefont {M~T}\ \bibnamefont {Wharmby}}, \bibinfo {author}
  {\bibfnamefont {J}~\bibnamefont {Chen}}, \bibinfo {author} {\bibfnamefont
  {A~L}\ \bibnamefont {Goodwin}}, \ and\ \bibinfo {author} {\bibfnamefont
  {M~S}\ \bibnamefont {Senn}},\ }\bibfield  {title} {\enquote {\bibinfo {title}
  {{Anharmonicity and scissoring modes in the negative thermal expansion
  materials ${\rm ScF}_{3}$ and ${\rm CaZrF}_{6}$}},}\ }\href {\doibase
  10.1103/PhysRevB.101.064306} {\bibfield  {journal} {\bibinfo  {journal}
  {Physical Review B}\ }\textbf {\bibinfo {volume} {101}},\ \bibinfo {pages}
  {064306} (\bibinfo {year} {2020})}\BibitemShut {NoStop}%
\bibitem [{Note2()}]{Note2}%
  \BibitemOpen
  \bibinfo {note} {The authors of reference \protect \rev@citealp {Li:2011dn}
  give a slightly misleading qualitative opinion on the fexibility of the
  ScF$_6$ octahedra on the basis of ab initio molecular dynamics simulations,
  because they constructed their atomic configuration with an \protect \textit
  {odd} number of unit cells along each direction. This choice automatically
  excludes all Rigid Unit Modes, and thus their simulation is unrealistic and
  their conclusions, albeit qualitative, are affected by this
  choice.}\BibitemShut {Stop}%
\bibitem [{\citenamefont {Fang}\ \emph {et~al.}(2014)\citenamefont {Fang},
  \citenamefont {Dove},\ and\ \citenamefont {Phillips}}]{Fang:2014cp}%
  \BibitemOpen
  \bibfield  {author} {\bibinfo {author} {\bibfnamefont {Hong}\ \bibnamefont
  {Fang}}, \bibinfo {author} {\bibfnamefont {Martin~T}\ \bibnamefont {Dove}}, \
  and\ \bibinfo {author} {\bibfnamefont {Anthony~E}\ \bibnamefont {Phillips}},\
  }\bibfield  {title} {\enquote {\bibinfo {title} {{Common origin of negative
  thermal expansion and other exotic properties in ceramic and hybrid
  materials.}}}\ }\href {\doibase 10.1103/PhysRevB.89.214103} {\bibfield
  {journal} {\bibinfo  {journal} {Physical Review B}\ }\textbf {\bibinfo
  {volume} {89}},\ \bibinfo {pages} {214103} (\bibinfo {year}
  {2014})}\BibitemShut {NoStop}%
\bibitem [{\citenamefont {Oba}\ \emph {et~al.}(2019)\citenamefont {Oba},
  \citenamefont {Tadano}, \citenamefont {Akashi},\ and\ \citenamefont
  {Tsuneyuki}}]{Oba:2019bi}%
  \BibitemOpen
  \bibfield  {author} {\bibinfo {author} {\bibfnamefont {Yusuke}\ \bibnamefont
  {Oba}}, \bibinfo {author} {\bibfnamefont {Terumasa}\ \bibnamefont {Tadano}},
  \bibinfo {author} {\bibfnamefont {Ryosuke}\ \bibnamefont {Akashi}}, \ and\
  \bibinfo {author} {\bibfnamefont {Shinji}\ \bibnamefont {Tsuneyuki}},\
  }\bibfield  {title} {\enquote {\bibinfo {title} {{First-principles study of
  phonon anharmonicity and negative thermal expansion in ${\rm ScF}_{3}$}},}\
  }\href {\doibase 10.1103/PhysRevMaterials.3.033601} {\bibfield  {journal}
  {\bibinfo  {journal} {Physical Review Materials}\ }\textbf {\bibinfo {volume}
  {3}},\ \bibinfo {pages} {033601} (\bibinfo {year} {2019})}\BibitemShut
  {NoStop}%
\bibitem [{\citenamefont {Handunkanda}\ \emph {et~al.}(2015)\citenamefont
  {Handunkanda}, \citenamefont {Curry}, \citenamefont {Voronov}, \citenamefont
  {Said}, \citenamefont {Guzm{\'a}n-Verri}, \citenamefont {Brierley},
  \citenamefont {Littlewood},\ and\ \citenamefont
  {Hancock}}]{Handunkanda:2015dc}%
  \BibitemOpen
  \bibfield  {author} {\bibinfo {author} {\bibfnamefont {Sahan~U}\ \bibnamefont
  {Handunkanda}}, \bibinfo {author} {\bibfnamefont {Erin~B}\ \bibnamefont
  {Curry}}, \bibinfo {author} {\bibfnamefont {Vladimir}\ \bibnamefont
  {Voronov}}, \bibinfo {author} {\bibfnamefont {Ayman~H}\ \bibnamefont {Said}},
  \bibinfo {author} {\bibfnamefont {Gian~G}\ \bibnamefont {Guzm{\'a}n-Verri}},
  \bibinfo {author} {\bibfnamefont {Richard~T}\ \bibnamefont {Brierley}},
  \bibinfo {author} {\bibfnamefont {Peter~B}\ \bibnamefont {Littlewood}}, \
  and\ \bibinfo {author} {\bibfnamefont {Jason~N}\ \bibnamefont {Hancock}},\
  }\bibfield  {title} {\enquote {\bibinfo {title} {{Large isotropic negative
  thermal expansion above a structural quantum phase transition.}}}\ }\href
  {\doibase 10.1103/PhysRevB.92.134101} {\bibfield  {journal} {\bibinfo
  {journal} {Physical Review B}\ }\textbf {\bibinfo {volume} {92}},\ \bibinfo
  {pages} {134101--6} (\bibinfo {year} {2015})}\BibitemShut {NoStop}%
\bibitem [{\citenamefont {Handunkanda}\ \emph {et~al.}(2016)\citenamefont
  {Handunkanda}, \citenamefont {Occhialini}, \citenamefont {Said},\ and\
  \citenamefont {Hancock}}]{Handunkanda:2016et}%
  \BibitemOpen
  \bibfield  {author} {\bibinfo {author} {\bibfnamefont {Sahan~U}\ \bibnamefont
  {Handunkanda}}, \bibinfo {author} {\bibfnamefont {Connor~A}\ \bibnamefont
  {Occhialini}}, \bibinfo {author} {\bibfnamefont {Ayman~H}\ \bibnamefont
  {Said}}, \ and\ \bibinfo {author} {\bibfnamefont {Jason~N}\ \bibnamefont
  {Hancock}},\ }\bibfield  {title} {\enquote {\bibinfo {title}
  {{Two-dimensional nanoscale correlations in the strong negative thermal
  expansion material ScF$_3$}},}\ }\href@noop {} {\bibfield  {journal}
  {\bibinfo  {journal} {Physical Review B}\ }\textbf {\bibinfo {volume} {94}},\
  \bibinfo {pages} {214102--6} (\bibinfo {year} {2016})}\BibitemShut {NoStop}%
\bibitem [{\citenamefont {Giddy}\ \emph {et~al.}(1993)\citenamefont {Giddy},
  \citenamefont {Dove}, \citenamefont {Pawley},\ and\ \citenamefont
  {Heine}}]{Giddy:1993ue}%
  \BibitemOpen
  \bibfield  {author} {\bibinfo {author} {\bibfnamefont {A~P}\ \bibnamefont
  {Giddy}}, \bibinfo {author} {\bibfnamefont {M~T}\ \bibnamefont {Dove}},
  \bibinfo {author} {\bibfnamefont {G~S}\ \bibnamefont {Pawley}}, \ and\
  \bibinfo {author} {\bibfnamefont {V}~\bibnamefont {Heine}},\ }\bibfield
  {title} {\enquote {\bibinfo {title} {{The determination of rigid-unit modes
  as potential soft modes for displacive phase transitions in framework crystal
  structures.}}}\ }\href {\doibase 10.1107/S0108767393002545} {\bibfield
  {journal} {\bibinfo  {journal} {Acta Crystallographica Section A: Foundations
  of Crystallography}\ }\textbf {\bibinfo {volume} {49}},\ \bibinfo {pages}
  {697--703} (\bibinfo {year} {1993})}\BibitemShut {NoStop}%
\bibitem [{\citenamefont {Hammonds}\ \emph {et~al.}(1996)\citenamefont
  {Hammonds}, \citenamefont {Dove}, \citenamefont {Giddy}, \citenamefont
  {Heine},\ and\ \citenamefont {Winkler}}]{Hammonds:1996wy}%
  \BibitemOpen
  \bibfield  {author} {\bibinfo {author} {\bibfnamefont {Kenton~D}\
  \bibnamefont {Hammonds}}, \bibinfo {author} {\bibfnamefont {Martin~T}\
  \bibnamefont {Dove}}, \bibinfo {author} {\bibfnamefont {Andrew~P}\
  \bibnamefont {Giddy}}, \bibinfo {author} {\bibfnamefont {Volker}\
  \bibnamefont {Heine}}, \ and\ \bibinfo {author} {\bibfnamefont {Bjoern}\
  \bibnamefont {Winkler}},\ }\bibfield  {title} {\enquote {\bibinfo {title}
  {{Rigid-unit phonon modes and structural phase transitions in framework
  silicates}},}\ }\href {\doibase 10.2138/am-1996-9-1003} {\bibfield  {journal}
  {\bibinfo  {journal} {American Mineralogist}\ }\textbf {\bibinfo {volume}
  {81}},\ \bibinfo {pages} {1057--1079} (\bibinfo {year} {1996})}\BibitemShut
  {NoStop}%
\bibitem [{\citenamefont {Heine}\ \emph {et~al.}(1999)\citenamefont {Heine},
  \citenamefont {Welche},\ and\ \citenamefont {Dove}}]{Heine:1999vk}%
  \BibitemOpen
  \bibfield  {author} {\bibinfo {author} {\bibfnamefont {Volker}\ \bibnamefont
  {Heine}}, \bibinfo {author} {\bibfnamefont {Patrick R~L}\ \bibnamefont
  {Welche}}, \ and\ \bibinfo {author} {\bibfnamefont {Martin~T}\ \bibnamefont
  {Dove}},\ }\bibfield  {title} {\enquote {\bibinfo {title} {{Geometrical
  Origin and Theory of Negative Thermal Expansion in Framework Structures}},}\
  }\href {\doibase 10.1111/j.1151-2916.1999.tb02001.x} {\bibfield  {journal}
  {\bibinfo  {journal} {Journal of the American Ceramic Society}\ }\textbf
  {\bibinfo {volume} {82}},\ \bibinfo {pages} {1793--1802} (\bibinfo {year}
  {1999})}\BibitemShut {NoStop}%
\bibitem [{\citenamefont {Dove}(2019)}]{Dove:2019dl}%
  \BibitemOpen
  \bibfield  {author} {\bibinfo {author} {\bibfnamefont {Martin~T}\
  \bibnamefont {Dove}},\ }\bibfield  {title} {\enquote {\bibinfo {title}
  {{Flexibility of network materials and the Rigid Unit Mode model: a personal
  perspective}},}\ }\href {\doibase 10.1098/rsta.2018.0222} {\bibfield
  {journal} {\bibinfo  {journal} {Philosophical Transactions of the Royal
  Society A: Mathematical, Physical and Engineering Sciences}\ }\textbf
  {\bibinfo {volume} {377}},\ \bibinfo {pages} {20180222--18} (\bibinfo {year}
  {2019})}\BibitemShut {NoStop}%
\bibitem [{\citenamefont {Stirling}(1972)}]{Stirling:1972ul}%
  \BibitemOpen
  \bibfield  {author} {\bibinfo {author} {\bibfnamefont {W~G}\ \bibnamefont
  {Stirling}},\ }\bibfield  {title} {\enquote {\bibinfo {title} {{Neutron
  inelastic scattering study of the lattice dynamics of strontium titanate:
  harmonic models.}}}\ }\href {\doibase 10.1088/0022-3719/5/19/005} {\bibfield
  {journal} {\bibinfo  {journal} {Journal of Physics C: Solid State Physics}\
  ,\ \bibinfo {pages} {2711--2730}} (\bibinfo {year} {1972})}\BibitemShut
  {NoStop}%
\bibitem [{Note3()}]{Note3}%
  \BibitemOpen
  \bibinfo {note} {It is worth remarking about the origin of the stiffness in
  the structural polyhedra in general and more specifically in the case of the
  ScF$_6$ octahedra in ScF$_3$. Much of the original literature on RUMs
  concerned silica and silicates, which are conventionally considered to have
  strong covalent bonds defining the shape and stiffness of the structural
  SiO$_4$ tetrahedra. However, it is important to understand that the rigidity
  of structural polyhedra do not rely on covalent bonding, because tension
  within the polyhedra in a system where the bonding is more ionic in nature
  can arise from mutual repulsions between the vertex ions because of size
  effects and Coulomb interactions. This is pertinent for ScF$_3$. The Sc--F
  bond is relatively strong, as evidenced by the large value of the Sc--F
  stretching frequency in the phonon dispersion curves \cite
  {Li:2011dn,Handunkanda:2015dc,Oba:2019bi}, indicating some degree of covalent
  bonding, but it is more likely that the ionic size and electrostatic
  interactions will be most important in the F--Sc--F bond-bending
  forces.}\BibitemShut {Stop}%
\bibitem [{POL(2019)}]{POLARIStech}%
  \BibitemOpen
  \href {https://www.isis.stfc.ac.uk/Pages/Polaris-technical-information.aspx}
  {\enquote {\bibinfo {title} {Polaris technical information, {STFC ISIS
  Neutron and Muon Source},
  https://www.isis.stfc.ac.uk/pages/polaris-technical-information.aspx},}\ }
  (\bibinfo {year} {2019})\BibitemShut {NoStop}%
\bibitem [{\citenamefont {Larson}\ and\ \citenamefont
  {Von~Dreele}(2004)}]{Larson:2004wv}%
  \BibitemOpen
  \bibfield  {author} {\bibinfo {author} {\bibfnamefont {A~C}\ \bibnamefont
  {Larson}}\ and\ \bibinfo {author} {\bibfnamefont {R~B}\ \bibnamefont
  {Von~Dreele}},\ }\href@noop {} {\emph {\bibinfo {title} {{General Structure
  Analysis System (GSAS)}}}},\ \bibinfo {type} {Tech. Rep.}\ \bibinfo {number}
  {LAUR 86-748}\ (\bibinfo  {institution} {Los Alamos National Laboratory},\
  \bibinfo {year} {2004})\BibitemShut {NoStop}%
\bibitem [{\citenamefont {Toby}(2001)}]{gsasgui}%
  \BibitemOpen
  \bibfield  {author} {\bibinfo {author} {\bibfnamefont {Brian~H}\ \bibnamefont
  {Toby}},\ }\bibfield  {title} {\enquote {\bibinfo {title} {{EXPGUI, a
  graphical user interface for GSAS}},}\ }\href {\doibase
  10.1107/S0021889801002242} {\bibfield  {journal} {\bibinfo  {journal}
  {Journal of Applied Crystallography}\ }\textbf {\bibinfo {volume} {34}},\
  \bibinfo {pages} {210--213} (\bibinfo {year} {2001})}\BibitemShut {NoStop}%
\bibitem [{\citenamefont {Arnold}\ \emph {et~al.}(2014)\citenamefont {Arnold},
  \citenamefont {Bilheux}, \citenamefont {Borreguero}, \citenamefont {Buts},
  \citenamefont {Campbell}, \citenamefont {Chapon}, \citenamefont {Doucet},
  \citenamefont {Draper}, \citenamefont {Leal}, \citenamefont {Gigg},
  \citenamefont {Lynch}, \citenamefont {Markvardsen}, \citenamefont
  {Mikkelson}, \citenamefont {Mikkelson}, \citenamefont {Miller}, \citenamefont
  {Palmen}, \citenamefont {Parker}, \citenamefont {Passos}, \citenamefont
  {Perring}, \citenamefont {Peterson}, \citenamefont {Ren}, \citenamefont
  {Reuter}, \citenamefont {Savici}, \citenamefont {Taylor}, \citenamefont
  {Taylor}, \citenamefont {Tolchenov}, \citenamefont {Zhou},\ and\
  \citenamefont {Zikovsky}}]{Arnold:2014iy}%
  \BibitemOpen
  \bibfield  {author} {\bibinfo {author} {\bibfnamefont {O}~\bibnamefont
  {Arnold}}, \bibinfo {author} {\bibfnamefont {J~C}\ \bibnamefont {Bilheux}},
  \bibinfo {author} {\bibfnamefont {J~M}\ \bibnamefont {Borreguero}}, \bibinfo
  {author} {\bibfnamefont {A}~\bibnamefont {Buts}}, \bibinfo {author}
  {\bibfnamefont {S~I}\ \bibnamefont {Campbell}}, \bibinfo {author}
  {\bibfnamefont {L}~\bibnamefont {Chapon}}, \bibinfo {author} {\bibfnamefont
  {M}~\bibnamefont {Doucet}}, \bibinfo {author} {\bibfnamefont {N}~\bibnamefont
  {Draper}}, \bibinfo {author} {\bibfnamefont {R~Ferraz}\ \bibnamefont {Leal}},
  \bibinfo {author} {\bibfnamefont {M~A}\ \bibnamefont {Gigg}}, \bibinfo
  {author} {\bibfnamefont {V~E}\ \bibnamefont {Lynch}}, \bibinfo {author}
  {\bibfnamefont {A}~\bibnamefont {Markvardsen}}, \bibinfo {author}
  {\bibfnamefont {D~J}\ \bibnamefont {Mikkelson}}, \bibinfo {author}
  {\bibfnamefont {R~L}\ \bibnamefont {Mikkelson}}, \bibinfo {author}
  {\bibfnamefont {R}~\bibnamefont {Miller}}, \bibinfo {author} {\bibfnamefont
  {K}~\bibnamefont {Palmen}}, \bibinfo {author} {\bibfnamefont {P}~\bibnamefont
  {Parker}}, \bibinfo {author} {\bibfnamefont {G}~\bibnamefont {Passos}},
  \bibinfo {author} {\bibfnamefont {T~G}\ \bibnamefont {Perring}}, \bibinfo
  {author} {\bibfnamefont {P~F}\ \bibnamefont {Peterson}}, \bibinfo {author}
  {\bibfnamefont {S}~\bibnamefont {Ren}}, \bibinfo {author} {\bibfnamefont
  {M~A}\ \bibnamefont {Reuter}}, \bibinfo {author} {\bibfnamefont {A~T}\
  \bibnamefont {Savici}}, \bibinfo {author} {\bibfnamefont {J~W}\ \bibnamefont
  {Taylor}}, \bibinfo {author} {\bibfnamefont {R~J}\ \bibnamefont {Taylor}},
  \bibinfo {author} {\bibfnamefont {R}~\bibnamefont {Tolchenov}}, \bibinfo
  {author} {\bibfnamefont {W}~\bibnamefont {Zhou}}, \ and\ \bibinfo {author}
  {\bibfnamefont {J}~\bibnamefont {Zikovsky}},\ }\bibfield  {title} {\enquote
  {\bibinfo {title} {{Mantid---Data analysis and visualization package for
  neutron scattering and $\mu$ SR experiments}},}\ }\href@noop {} {\bibfield
  {journal} {\bibinfo  {journal} {Nuclear Inst. and Methods in Physics
  Research, A}\ }\textbf {\bibinfo {volume} {764}},\ \bibinfo {pages}
  {156--166} (\bibinfo {year} {2014})}\BibitemShut {NoStop}%
\bibitem [{\citenamefont {Tucker}\ \emph
  {et~al.}(2007{\natexlab{b}})\citenamefont {Tucker}, \citenamefont {Keen},
  \citenamefont {Dove}, \citenamefont {Goodwin},\ and\ \citenamefont
  {Hui}}]{Tucker:2007eh}%
  \BibitemOpen
  \bibfield  {author} {\bibinfo {author} {\bibfnamefont {Matthew~G}\
  \bibnamefont {Tucker}}, \bibinfo {author} {\bibfnamefont {David~A}\
  \bibnamefont {Keen}}, \bibinfo {author} {\bibfnamefont {Martin~T}\
  \bibnamefont {Dove}}, \bibinfo {author} {\bibfnamefont {Andrew~L}\
  \bibnamefont {Goodwin}}, \ and\ \bibinfo {author} {\bibfnamefont {Qun}\
  \bibnamefont {Hui}},\ }\bibfield  {title} {\enquote {\bibinfo {title}
  {{RMCProfile: reverse Monte Carlo for polycrystalline materials.}}}\ }\href
  {\doibase 10.1088/0953-8984/19/33/335218} {\bibfield  {journal} {\bibinfo
  {journal} {Journal of Physics: Condensed Matter}\ }\textbf {\bibinfo {volume}
  {19}},\ \bibinfo {pages} {335218--16} (\bibinfo {year}
  {2007}{\natexlab{b}})}\BibitemShut {NoStop}%
\bibitem [{\citenamefont {Soper}(2012)}]{Soper:2012vs}%
  \BibitemOpen
  \bibfield  {author} {\bibinfo {author} {\bibfnamefont {Alan~K}\ \bibnamefont
  {Soper}},\ }\href
  {https://www.isis.stfc.ac.uk/OtherFiles/Disordered%20Materials/Gudrun-Manual-2017-10.pdf}
  {\emph {\bibinfo {title} {{GudrunN and GudrunX}}}},\ \bibinfo {type} {Tech.
  Rep.}\ \bibinfo {number} {RAL-TR 13}\ (\bibinfo  {institution} {Rutherford
  Appleton Laboratory},\ \bibinfo {year} {2012})\BibitemShut {NoStop}%
\bibitem [{\citenamefont {Todorov}\ \emph {et~al.}(2006)\citenamefont
  {Todorov}, \citenamefont {Smith}, \citenamefont {Trachenko},\ and\
  \citenamefont {Dove}}]{Todorov:2006ee}%
  \BibitemOpen
  \bibfield  {author} {\bibinfo {author} {\bibfnamefont {Ilian~T}\ \bibnamefont
  {Todorov}}, \bibinfo {author} {\bibfnamefont {William}\ \bibnamefont
  {Smith}}, \bibinfo {author} {\bibfnamefont {Kostya}\ \bibnamefont
  {Trachenko}}, \ and\ \bibinfo {author} {\bibfnamefont {Martin~T}\
  \bibnamefont {Dove}},\ }\bibfield  {title} {\enquote {\bibinfo {title}
  {{DL\_POLY\_3: new dimensions in molecular dynamics simulations via massive
  parallelism}},}\ }\href {\doibase 10.1039/b517931a} {\bibfield  {journal}
  {\bibinfo  {journal} {Journal of Materials Chemistry}\ }\textbf {\bibinfo
  {volume} {16}},\ \bibinfo {pages} {1911--8} (\bibinfo {year}
  {2006})}\BibitemShut {NoStop}%
\bibitem [{\citenamefont {Gale}(1997)}]{Gale:1997iq}%
  \BibitemOpen
  \bibfield  {author} {\bibinfo {author} {\bibfnamefont {Julian~D}\
  \bibnamefont {Gale}},\ }\bibfield  {title} {\enquote {\bibinfo {title}
  {{GULP: A computer program for the symmetry-adapted simulation of solids}},}\
  }\href {\doibase 10.1039/a606455h} {\bibfield  {journal} {\bibinfo  {journal}
  {Journal of the Chemical Society, Faraday Transactions}\ }\textbf {\bibinfo
  {volume} {93}},\ \bibinfo {pages} {629--637} (\bibinfo {year}
  {1997})}\BibitemShut {NoStop}%
\bibitem [{\citenamefont {Gale}\ and\ \citenamefont
  {Rohl}(2003)}]{Gale:2003eo}%
  \BibitemOpen
  \bibfield  {author} {\bibinfo {author} {\bibfnamefont {Julian~D}\
  \bibnamefont {Gale}}\ and\ \bibinfo {author} {\bibfnamefont {Andrew~L}\
  \bibnamefont {Rohl}},\ }\bibfield  {title} {\enquote {\bibinfo {title} {{The
  General Utility Lattice Program (GULP).}}}\ }\href {\doibase
  10.1080/0892702031000104887} {\bibfield  {journal} {\bibinfo  {journal}
  {Molecular Simulation}\ }\textbf {\bibinfo {volume} {29}},\ \bibinfo {pages}
  {291--341} (\bibinfo {year} {2003})}\BibitemShut {NoStop}%
\bibitem [{\citenamefont {Tucker}\ \emph {et~al.}(2001)\citenamefont {Tucker},
  \citenamefont {Dove},\ and\ \citenamefont {Keen}}]{Tucker:2001fh}%
  \BibitemOpen
  \bibfield  {author} {\bibinfo {author} {\bibfnamefont {Matthew~G}\
  \bibnamefont {Tucker}}, \bibinfo {author} {\bibfnamefont {Martin~T}\
  \bibnamefont {Dove}}, \ and\ \bibinfo {author} {\bibfnamefont {David~A}\
  \bibnamefont {Keen}},\ }\bibfield  {title} {\enquote {\bibinfo {title}
  {{Application of the reverse Monte Carlo method to crystalline materials}},}\
  }\href {\doibase 10.1107/S002188980100930X} {\bibfield  {journal} {\bibinfo
  {journal} {Journal of Applied Crystallography}\ }\textbf {\bibinfo {volume}
  {34}},\ \bibinfo {pages} {630--638} (\bibinfo {year} {2001})}\BibitemShut
  {NoStop}%
\bibitem [{\citenamefont {Dove}\ \emph {et~al.}(2002)\citenamefont {Dove},
  \citenamefont {Tucker},\ and\ \citenamefont {Keen}}]{Dove:2002ed}%
  \BibitemOpen
  \bibfield  {author} {\bibinfo {author} {\bibfnamefont {Martin~T}\
  \bibnamefont {Dove}}, \bibinfo {author} {\bibfnamefont {Matthew~G}\
  \bibnamefont {Tucker}}, \ and\ \bibinfo {author} {\bibfnamefont {David~A}\
  \bibnamefont {Keen}},\ }\bibfield  {title} {\enquote {\bibinfo {title}
  {{Neutron total scattering method: simultaneous determination of long-range
  and short-range order in disordered materials.}}}\ }\href {\doibase
  10.1127/0935-1221/2002/0014-0331} {\bibfield  {journal} {\bibinfo  {journal}
  {European Journal of Mineralogy}\ }\textbf {\bibinfo {volume} {14}},\
  \bibinfo {pages} {331--348} (\bibinfo {year} {2002})}\BibitemShut {NoStop}%
\bibitem [{\citenamefont {Keen}\ \emph {et~al.}(2005)\citenamefont {Keen},
  \citenamefont {Tucker},\ and\ \citenamefont {Dove}}]{Keen:2005dd}%
  \BibitemOpen
  \bibfield  {author} {\bibinfo {author} {\bibfnamefont {D~A}\ \bibnamefont
  {Keen}}, \bibinfo {author} {\bibfnamefont {M~G}\ \bibnamefont {Tucker}}, \
  and\ \bibinfo {author} {\bibfnamefont {M~T}\ \bibnamefont {Dove}},\
  }\bibfield  {title} {\enquote {\bibinfo {title} {{Reverse Monte Carlo
  modelling of crystalline disorder}},}\ }\href {\doibase
  10.1088/0953-8984/17/5/002} {\bibfield  {journal} {\bibinfo  {journal}
  {Journal of Physics: Condensed Matter}\ }\textbf {\bibinfo {volume} {17}},\
  \bibinfo {pages} {S15--S22} (\bibinfo {year} {2005})}\BibitemShut {NoStop}%
\bibitem [{Note4()}]{Note4}%
  \BibitemOpen
  \bibinfo {note} {The recent study of Wendt et al \cite {Wendt:2019it} reports
  two anomalous findings, first that the mean F--F distance shrinks with
  increasing temperature, and second that the integrated area of the second
  F--F peak in the PDF decreases with temperature. In both cases these are
  contrary to what we have found. These two anomalous results may reflect the
  limitations of analysis of the PDF alone, without the support of a model. As
  the F--F peak broadens with temperature, the data become increasingly noisy,
  and the pair density within any fixed $r$ range will necessarily decrease.
  Furthermore, the analysis may also be affected by artefacts associated with
  untreated truncation effects at the maximum observable value of $Q$. Such
  problems highlight the value of making use of modelling approaches such as
  RMC, where the atomic model imposes reasonable physical constraints on the
  data analysis. The atomistic configuration RMC provides means that there is
  no ambiguity in assigning pair density where peaks are very broad or overlap,
  such as the 4~\r A\ peak here, which encompasses both Sc--Sc and F$\protect
  \cdots $F pairs. Figures S7--S9 in the Supplementary Information show the
  RMC-derived decomposition of the PDF into contributions from individual
  atomic pairs.}\BibitemShut {Stop}%
\bibitem [{\citenamefont {Fang}\ \emph
  {et~al.}(2013{\natexlab{b}})\citenamefont {Fang}, \citenamefont {Dove},
  \citenamefont {Rimmer},\ and\ \citenamefont {Misquitta}}]{Fang:2013ji}%
  \BibitemOpen
  \bibfield  {author} {\bibinfo {author} {\bibfnamefont {Hong}\ \bibnamefont
  {Fang}}, \bibinfo {author} {\bibfnamefont {Martin~T}\ \bibnamefont {Dove}},
  \bibinfo {author} {\bibfnamefont {Leila H~N}\ \bibnamefont {Rimmer}}, \ and\
  \bibinfo {author} {\bibfnamefont {Alston~J}\ \bibnamefont {Misquitta}},\
  }\bibfield  {title} {\enquote {\bibinfo {title} {{Simulation study of
  pressure and temperature dependence of the negative thermal expansion in
  Zn(CN)$_2$.}}}\ }\href {\doibase 10.1103/PhysRevB.88.104306} {\bibfield
  {journal} {\bibinfo  {journal} {Physical Review B}\ }\textbf {\bibinfo
  {volume} {88}},\ \bibinfo {pages} {104306} (\bibinfo {year}
  {2013}{\natexlab{b}})}\BibitemShut {NoStop}%
\bibitem [{\citenamefont {Aleksandrov}\ \emph {et~al.}(2002)\citenamefont
  {Aleksandrov}, \citenamefont {Voronov}, \citenamefont {Vtyurin},
  \citenamefont {Goryainov}, \citenamefont {Zamkova}, \citenamefont {Zinenko},\
  and\ \citenamefont {Krylov}}]{Aleksandrov:2002ut}%
  \BibitemOpen
  \bibfield  {author} {\bibinfo {author} {\bibfnamefont {K~S}\ \bibnamefont
  {Aleksandrov}}, \bibinfo {author} {\bibfnamefont {V~N}\ \bibnamefont
  {Voronov}}, \bibinfo {author} {\bibfnamefont {A~N}\ \bibnamefont {Vtyurin}},
  \bibinfo {author} {\bibfnamefont {S~V}\ \bibnamefont {Goryainov}}, \bibinfo
  {author} {\bibfnamefont {N~G}\ \bibnamefont {Zamkova}}, \bibinfo {author}
  {\bibfnamefont {V~I}\ \bibnamefont {Zinenko}}, \ and\ \bibinfo {author}
  {\bibfnamefont {A~S}\ \bibnamefont {Krylov}},\ }\bibfield  {title} {\enquote
  {\bibinfo {title} {{Lattice dynamics and hydrostatic-pressure-induced phase
  transitions in ScF$_3$.}}}\ }\href {\doibase 10.1134/1.1484991} {\bibfield
  {journal} {\bibinfo  {journal} {Journal of Experimental and Theoretical
  Physics}\ }\textbf {\bibinfo {volume} {94}},\ \bibinfo {pages} {977--984}
  (\bibinfo {year} {2002})}\BibitemShut {NoStop}%
\bibitem [{\citenamefont {Wells}\ \emph
  {et~al.}(2002{\natexlab{a}})\citenamefont {Wells}, \citenamefont {Dove},\
  and\ \citenamefont {Tucker}}]{Wells:2002ty}%
  \BibitemOpen
  \bibfield  {author} {\bibinfo {author} {\bibfnamefont {Stephen~A}\
  \bibnamefont {Wells}}, \bibinfo {author} {\bibfnamefont {Martin~T}\
  \bibnamefont {Dove}}, \ and\ \bibinfo {author} {\bibfnamefont {Matthew~G}\
  \bibnamefont {Tucker}},\ }\bibfield  {title} {\enquote {\bibinfo {title}
  {{Finding best-fit polyhedral rotations with geometric algebra}},}\ }\href
  {\doibase 10.1088/0953-8984/14/17/327} {\bibfield  {journal} {\bibinfo
  {journal} {Journal of Physics: Condensed Matter}\ }\textbf {\bibinfo {volume}
  {14}},\ \bibinfo {pages} {4567--4584} (\bibinfo {year}
  {2002}{\natexlab{a}})}\BibitemShut {NoStop}%
\bibitem [{\citenamefont {Wells}\ \emph
  {et~al.}(2002{\natexlab{b}})\citenamefont {Wells}, \citenamefont {Dove},
  \citenamefont {Tucker},\ and\ \citenamefont {Trachenko}}]{Wells:2002tq}%
  \BibitemOpen
  \bibfield  {author} {\bibinfo {author} {\bibfnamefont {Stephen~A}\
  \bibnamefont {Wells}}, \bibinfo {author} {\bibfnamefont {Martin~T}\
  \bibnamefont {Dove}}, \bibinfo {author} {\bibfnamefont {Matthew~G}\
  \bibnamefont {Tucker}}, \ and\ \bibinfo {author} {\bibfnamefont {Kostya}\
  \bibnamefont {Trachenko}},\ }\bibfield  {title} {\enquote {\bibinfo {title}
  {{Real-space rigid-unit-mode analysis of dynamic disorder in quartz,
  cristobalite and amorphous silica.}}}\ }\href {\doibase
  10.1088/0953-8984/14/18/302} {\bibfield  {journal} {\bibinfo  {journal}
  {Journal of Physics: Condensed Matter}\ }\textbf {\bibinfo {volume} {14}},\
  \bibinfo {pages} {4645--4657} (\bibinfo {year}
  {2002}{\natexlab{b}})}\BibitemShut {NoStop}%
\bibitem [{\citenamefont {Wells}\ \emph {et~al.}(2004)\citenamefont {Wells},
  \citenamefont {Dove},\ and\ \citenamefont {Tucker}}]{Wells:2004en}%
  \BibitemOpen
  \bibfield  {author} {\bibinfo {author} {\bibfnamefont {Stephen~A}\
  \bibnamefont {Wells}}, \bibinfo {author} {\bibfnamefont {Martin~T}\
  \bibnamefont {Dove}}, \ and\ \bibinfo {author} {\bibfnamefont {Matthew~G}\
  \bibnamefont {Tucker}},\ }\bibfield  {title} {\enquote {\bibinfo {title}
  {{Reverse Monte Carlo with geometric analysis {\textendash} RMC+GA}},}\
  }\href {\doibase 10.1107/S0021889804008957} {\bibfield  {journal} {\bibinfo
  {journal} {Journal of Applied Crystallography}\ }\textbf {\bibinfo {volume}
  {37}},\ \bibinfo {pages} {536--544} (\bibinfo {year} {2004})}\BibitemShut
  {NoStop}%
\bibitem [{Note5()}]{Note5}%
  \BibitemOpen
  \bibinfo {note} {In Figure \ref {fig:angles} the sizes of the fluctuations of
  the ScF$_6$ orientations and F--Sc--F angles are similar, but in Figure \ref
  {fig:gasp} the GASP analysis suggest that the motions of the F atoms
  associated with bond bending are rather larger than from rotations of the
  octahedra. These results are actually consistent with each other: the
  contribution to $M$ from the octahedral rotation will come from three axes,
  but there are 12 bending angles associated with deformation of the
  octahedra.}\BibitemShut {Stop}%
\bibitem [{\citenamefont {Hui}\ \emph {et~al.}(2005)\citenamefont {Hui},
  \citenamefont {Tucker}, \citenamefont {Dove}, \citenamefont {Wells},\ and\
  \citenamefont {Keen}}]{Hui:2005bg}%
  \BibitemOpen
  \bibfield  {author} {\bibinfo {author} {\bibfnamefont {Qun}\ \bibnamefont
  {Hui}}, \bibinfo {author} {\bibfnamefont {Matthew~G}\ \bibnamefont {Tucker}},
  \bibinfo {author} {\bibfnamefont {Martin~T}\ \bibnamefont {Dove}}, \bibinfo
  {author} {\bibfnamefont {Stephen~A}\ \bibnamefont {Wells}}, \ and\ \bibinfo
  {author} {\bibfnamefont {David~A}\ \bibnamefont {Keen}},\ }\bibfield  {title}
  {\enquote {\bibinfo {title} {{Total scattering and reverse Monte Carlo study
  of the 105 K displacive phase transition in strontium titanate}},}\ }\href
  {\doibase 10.1088/0953-8984/17/5/012} {\bibfield  {journal} {\bibinfo
  {journal} {Journal of Physics: Condensed Matter}\ }\textbf {\bibinfo {volume}
  {17}},\ \bibinfo {pages} {S111--S124} (\bibinfo {year} {2005})}\BibitemShut
  {NoStop}%
\bibitem [{Note6()}]{Note6}%
  \BibitemOpen
  \bibinfo {note} {But it would be wrong to say that it \protect \textit {is}
  the limiting case, because the model, being a simplified model, is not an
  exact representation of the actual material. In particular, the division
  between the three types of motion is not exactly the same. What we are
  demonstrating is that even in the limit of fairly flexible octahedra the GASP
  method will still identify whole-body rotations.}\BibitemShut {Stop}%
\bibitem [{Note7()}]{Note7}%
  \BibitemOpen
  \bibinfo {note} {There are two important points to highlight in the two
  diagrams in Figure \ref {fig:angles_and_gasp}. The first is that the angular
  fluctuations in (a) increase linearly with temperature, in line with
  expectations from classical quasi-harmonic lattice dynamics. Second is that
  the distribution of atomic motion over the three components in (b) is
  more-or-less constant with temperature, reflecting phonon eigenvectors and
  again being fully consistent with classical quasi-harmonic lattice dynamics.
  In particular, although thermal motion is large at the higher temperature, it
  does not lead to any breakdown of the classical picture, contrary to the
  opinion expressed in reference \protect \rev@citealp
  {Wendt:2019it}.}\BibitemShut {Stop}%
\bibitem [{Note8()}]{Note8}%
  \BibitemOpen
  \bibinfo {note} {We consider models in which the Sc--F bond is the sole rigid
  entity, with complete flexibility of the ScF$_6$ octahedra, as for example in
  reference \protect \rev@citealp {Wendt:2019it}, to be over-flexible and
  unable to provide useful insights, not least because such models will predict
  the wrong signs of the Gr{\"u}neisen parameters for many phonon modes. A
  similar situation was discussed with regard to NTE in Cu$_2$O in reference
  \protect \rev@citealp {Rimmer:2014dj}, where the true rigid entity is the
  linear O--Cu--O trimer, and models that consider only the O--Cu bonds as the
  rigid entities are too flexible.}\BibitemShut {Stop}%
\bibitem [{\citenamefont {Lazar}\ \emph {et~al.}(2015)\citenamefont {Lazar},
  \citenamefont {Bu{\v c}ko},\ and\ \citenamefont {Hafner}}]{Lazar:2015es}%
  \BibitemOpen
  \bibfield  {author} {\bibinfo {author} {\bibfnamefont {Petr}\ \bibnamefont
  {Lazar}}, \bibinfo {author} {\bibfnamefont {Tom{\'a}{\v s}}\ \bibnamefont
  {Bu{\v c}ko}}, \ and\ \bibinfo {author} {\bibfnamefont {J{\"u}rgen}\
  \bibnamefont {Hafner}},\ }\bibfield  {title} {\enquote {\bibinfo {title}
  {{Negative thermal expansion of ScF$_3$: Insights from density-functional
  molecular dynamics in the isothermal-isobaric ensemble.}}}\ }\href {\doibase
  10.1103/PhysRevB.92.224302} {\bibfield  {journal} {\bibinfo  {journal}
  {Physical Review B}\ }\textbf {\bibinfo {volume} {92}},\ \bibinfo {pages}
  {224302--6} (\bibinfo {year} {2015})}\BibitemShut {NoStop}%
\bibitem [{\citenamefont {Bocharov}\ \emph {et~al.}(2019)\citenamefont
  {Bocharov}, \citenamefont {Rafalskij}, \citenamefont {Krack}, \citenamefont
  {Putnina},\ and\ \citenamefont {Kuzmin}}]{Bocharov:2019kj}%
  \BibitemOpen
  \bibfield  {author} {\bibinfo {author} {\bibfnamefont {Dmitry}\ \bibnamefont
  {Bocharov}}, \bibinfo {author} {\bibfnamefont {Yuri}\ \bibnamefont
  {Rafalskij}}, \bibinfo {author} {\bibfnamefont {Matthias}\ \bibnamefont
  {Krack}}, \bibinfo {author} {\bibfnamefont {Mara}\ \bibnamefont {Putnina}}, \
  and\ \bibinfo {author} {\bibfnamefont {Alexei}\ \bibnamefont {Kuzmin}},\
  }\bibfield  {title} {\enquote {\bibinfo {title} {{Negative thermal expansion
  of ScF$_3$: first principles vs empirical molecular dynamics}},}\ }\href
  {\doibase 10.1088/1757-899X/503/1/012001} {\bibfield  {journal} {\bibinfo
  {journal} {IOP Conference Series: Materials Science and Engineering}\
  }\textbf {\bibinfo {volume} {503}},\ \bibinfo {pages} {012001--5} (\bibinfo
  {year} {2019})}\BibitemShut {NoStop}%
\bibitem [{\citenamefont {Bocharov}\ \emph {et~al.}(2020)\citenamefont
  {Bocharov}, \citenamefont {Krack}, \citenamefont {Rafalskij}, \citenamefont
  {Kuzmin},\ and\ \citenamefont {Purans}}]{Bocharov:2020iu}%
  \BibitemOpen
  \bibfield  {author} {\bibinfo {author} {\bibfnamefont {D}~\bibnamefont
  {Bocharov}}, \bibinfo {author} {\bibfnamefont {M}~\bibnamefont {Krack}},
  \bibinfo {author} {\bibfnamefont {Yu}~\bibnamefont {Rafalskij}}, \bibinfo
  {author} {\bibfnamefont {A}~\bibnamefont {Kuzmin}}, \ and\ \bibinfo {author}
  {\bibfnamefont {J}~\bibnamefont {Purans}},\ }\bibfield  {title} {\enquote
  {\bibinfo {title} {{Ab initio molecular dynamics simulations of negative
  thermal expansion in ScF$_3$: The effect of the supercell size}},}\ }\href
  {\doibase 10.1016/j.commatsci.2019.109198} {\bibfield  {journal} {\bibinfo
  {journal} {Computational Materials Science}\ }\textbf {\bibinfo {volume}
  {171}},\ \bibinfo {pages} {109198} (\bibinfo {year} {2020})}\BibitemShut
  {NoStop}%
\bibitem [{\citenamefont {van Roekeghem}\ \emph {et~al.}(2016)\citenamefont
  {van Roekeghem}, \citenamefont {Carrete},\ and\ \citenamefont
  {Mingo}}]{vanRoekeghem:2016kd}%
  \BibitemOpen
  \bibfield  {author} {\bibinfo {author} {\bibfnamefont {Ambroise}\
  \bibnamefont {van Roekeghem}}, \bibinfo {author} {\bibfnamefont {Jes{\'u}s}\
  \bibnamefont {Carrete}}, \ and\ \bibinfo {author} {\bibfnamefont {Natalio}\
  \bibnamefont {Mingo}},\ }\bibfield  {title} {\enquote {\bibinfo {title}
  {{Anomalous thermal conductivity and suppression of negative thermal
  expansion in ScF$_3$.}}}\ }\href {\doibase 10.1103/PhysRevB.94.020303}
  {\bibfield  {journal} {\bibinfo  {journal} {Physical Review B}\ }\textbf
  {\bibinfo {volume} {94}},\ \bibinfo {pages} {020303--5} (\bibinfo {year}
  {2016})}\BibitemShut {NoStop}%
\bibitem [{Note9()}]{Note9}%
  \BibitemOpen
  \bibinfo {note} {In common usage, the term \protect \emph {quasiharmonic}
  model considers the effect of changes in volume on the phonon frequencies
  through the consequent changes in harmonic force constants, giving new
  harmonic phonon frequencies as modified by the anharmonic coupling of force
  constants to volume. The term \protect \emph {renormalised phonon} model
  considers how a renormalised harmonic Hamiltonian can result from a
  mean-field treatment of the anharmonic interactions between phonons,
  typically focussing on the fourth-order interactions.}\BibitemShut {Stop}%
\bibitem [{\citenamefont {Dove}(2009)}]{Dove:2009bn}%
  \BibitemOpen
  \bibfield  {author} {\bibinfo {author} {\bibfnamefont {Martin~T}\
  \bibnamefont {Dove}},\ }\href {\doibase 10.1017/CBO9780511619885} {\emph
  {\bibinfo {title} {{Introduction to Lattice Dynamics}}}}\ (\bibinfo
  {publisher} {Cambridge University Press},\ \bibinfo {address} {Cambridge},\
  \bibinfo {year} {2009})\BibitemShut {NoStop}%
\bibitem [{Note10()}]{Note10}%
  \BibitemOpen
  \bibinfo {note} {Importantly, the reverse Monte Carlo method is indeed
  capable of generating such distributions in cases where they genuinely exist;
  one clear example is that of the cubic phase of cristobalite \cite
  {Tucker:2001vg}.}\BibitemShut {Stop}%
\bibitem [{\citenamefont {Du}\ \emph {et~al.}(2019)\citenamefont {Du},
  \citenamefont {Phillips}, \citenamefont {Arnold}, \citenamefont {Keen},
  \citenamefont {Tucker},\ and\ \citenamefont {Dove}}]{Du:2019ei}%
  \BibitemOpen
  \bibfield  {author} {\bibinfo {author} {\bibfnamefont {Juan}\ \bibnamefont
  {Du}}, \bibinfo {author} {\bibfnamefont {Anthony~E}\ \bibnamefont
  {Phillips}}, \bibinfo {author} {\bibfnamefont {Donna~C}\ \bibnamefont
  {Arnold}}, \bibinfo {author} {\bibfnamefont {David~A}\ \bibnamefont {Keen}},
  \bibinfo {author} {\bibfnamefont {Matthew~G}\ \bibnamefont {Tucker}}, \ and\
  \bibinfo {author} {\bibfnamefont {Martin~T}\ \bibnamefont {Dove}},\
  }\bibfield  {title} {\enquote {\bibinfo {title} {{Structural study of bismuth
  ferrite BiFeO$_3$ by neutron total scattering and the reverse Monte Carlo
  method}},}\ }\href {\doibase 10.1103/PhysRevB.100.104111} {\bibfield
  {journal} {\bibinfo  {journal} {Physical Review B}\ }\textbf {\bibinfo
  {volume} {100}},\ \bibinfo {pages} {104111} (\bibinfo {year}
  {2019})}\BibitemShut {NoStop}%
\bibitem [{\citenamefont {Dove}\ \emph {et~al.}(1992)\citenamefont {Dove},
  \citenamefont {Giddy},\ and\ \citenamefont {Heine}}]{Dove:1992db}%
  \BibitemOpen
  \bibfield  {author} {\bibinfo {author} {\bibfnamefont {Martin~T}\
  \bibnamefont {Dove}}, \bibinfo {author} {\bibfnamefont {A~P}\ \bibnamefont
  {Giddy}}, \ and\ \bibinfo {author} {\bibfnamefont {V}~\bibnamefont {Heine}},\
  }\bibfield  {title} {\enquote {\bibinfo {title} {{On the application of
  mean-field and landau theory to displacive phase transitions}},}\ }\href
  {\doibase 10.1080/00150199208016064} {\bibfield  {journal} {\bibinfo
  {journal} {Ferroelectrics}\ }\textbf {\bibinfo {volume} {136}},\ \bibinfo
  {pages} {33--49} (\bibinfo {year} {1992})}\BibitemShut {NoStop}%
\bibitem [{\citenamefont {Dove}\ \emph {et~al.}(1995)\citenamefont {Dove},
  \citenamefont {Heine},\ and\ \citenamefont {Hammonds}}]{Dove:1995cw}%
  \BibitemOpen
  \bibfield  {author} {\bibinfo {author} {\bibfnamefont {Martin~T}\
  \bibnamefont {Dove}}, \bibinfo {author} {\bibfnamefont {Volker}\ \bibnamefont
  {Heine}}, \ and\ \bibinfo {author} {\bibfnamefont {Kenton~D}\ \bibnamefont
  {Hammonds}},\ }\bibfield  {title} {\enquote {\bibinfo {title} {{Rigid unit
  modes in framework silicates}},}\ }\href {\doibase
  10.1180/minmag.1995.059.397.07} {\bibfield  {journal} {\bibinfo  {journal}
  {Mineralogical Magazine}\ }\textbf {\bibinfo {volume} {59}},\ \bibinfo
  {pages} {629--639} (\bibinfo {year} {1995})}\BibitemShut {NoStop}%
\bibitem [{\citenamefont {Dove}\ \emph {et~al.}(1996)\citenamefont {Dove},
  \citenamefont {Gambhir}, \citenamefont {Hammonds}, \citenamefont {Heine},\
  and\ \citenamefont {Pryde}}]{Dove:1996tc}%
  \BibitemOpen
  \bibfield  {author} {\bibinfo {author} {\bibfnamefont {Martin~T}\
  \bibnamefont {Dove}}, \bibinfo {author} {\bibfnamefont {Manoj}\ \bibnamefont
  {Gambhir}}, \bibinfo {author} {\bibfnamefont {Kenton~D}\ \bibnamefont
  {Hammonds}}, \bibinfo {author} {\bibfnamefont {Volker}\ \bibnamefont
  {Heine}}, \ and\ \bibinfo {author} {\bibfnamefont {Alexandra K~A}\
  \bibnamefont {Pryde}},\ }\bibfield  {title} {\enquote {\bibinfo {title}
  {{Distortions of framework structures.}}}\ }\href {\doibase
  10.1080/01411599608242398} {\bibfield  {journal} {\bibinfo  {journal} {Phase
  Transitions}\ }\textbf {\bibinfo {volume} {58}},\ \bibinfo {pages} {121--143}
  (\bibinfo {year} {1996})}\BibitemShut {NoStop}%
\bibitem [{\citenamefont {Gambhir}\ \emph {et~al.}(1997)\citenamefont
  {Gambhir}, \citenamefont {Heine},\ and\ \citenamefont
  {Dove}}]{Gambhir:1997gv}%
  \BibitemOpen
  \bibfield  {author} {\bibinfo {author} {\bibfnamefont {Manoj}\ \bibnamefont
  {Gambhir}}, \bibinfo {author} {\bibfnamefont {Volker}\ \bibnamefont {Heine}},
  \ and\ \bibinfo {author} {\bibfnamefont {Martin~T}\ \bibnamefont {Dove}},\
  }\bibfield  {title} {\enquote {\bibinfo {title} {{A one-parameter model of a
  rigid-unit structure}},}\ }\href {\doibase 10.1080/01411599708223733}
  {\bibfield  {journal} {\bibinfo  {journal} {Phase Transitions}\ }\textbf
  {\bibinfo {volume} {61}},\ \bibinfo {pages} {125--139} (\bibinfo {year}
  {1997})}\BibitemShut {NoStop}%
\bibitem [{\citenamefont {Dove}\ \emph {et~al.}(1999)\citenamefont {Dove},
  \citenamefont {Gambhir},\ and\ \citenamefont {Heine}}]{Dove:1999uu}%
  \BibitemOpen
  \bibfield  {author} {\bibinfo {author} {\bibfnamefont {Martin~T}\
  \bibnamefont {Dove}}, \bibinfo {author} {\bibfnamefont {Manoj}\ \bibnamefont
  {Gambhir}}, \ and\ \bibinfo {author} {\bibfnamefont {Volker}\ \bibnamefont
  {Heine}},\ }\bibfield  {title} {\enquote {\bibinfo {title} {{Anatomy of a
  structural phase transition: theoretical analysis of the displacive phase
  transition in quartz and other silicates.}}}\ }\href {\doibase
  10.1007/s002690050194} {\bibfield  {journal} {\bibinfo  {journal} {Physics
  and Chemistry of Minerals}\ }\textbf {\bibinfo {volume} {26}},\ \bibinfo
  {pages} {344--353} (\bibinfo {year} {1999})}\BibitemShut {NoStop}%
\bibitem [{\citenamefont {Tkachenko}\ and\ \citenamefont
  {Zaliznyak}(2019)}]{Tkachenko:2020}%
  \BibitemOpen
  \bibfield  {author} {\bibinfo {author} {\bibfnamefont {Alexey~V}\
  \bibnamefont {Tkachenko}}\ and\ \bibinfo {author} {\bibfnamefont {Igor~A}\
  \bibnamefont {Zaliznyak}},\ }\href {https://arxiv.org/abs/1908.11643}
  {\enquote {\bibinfo {title} {Criticality, entropic elasticity, and negative
  thermal expansion of a coulomb floppy network: {ScF$_3$}-inspired theory for
  a class of ionic solids},}\ } (\bibinfo {year} {2019}),\ \bibinfo {note}
  {arXiV:1908.11643}\BibitemShut {NoStop}%
\bibitem [{\citenamefont {Wang}\ \emph {et~al.}(2015)\citenamefont {Wang},
  \citenamefont {Wang}, \citenamefont {Sun}, \citenamefont {Shi}, \citenamefont
  {Deng}, \citenamefont {Lu}, \citenamefont {Hu},\ and\ \citenamefont
  {Zhang}}]{Wang:2015eo}%
  \BibitemOpen
  \bibfield  {author} {\bibinfo {author} {\bibfnamefont {Lei}\ \bibnamefont
  {Wang}}, \bibinfo {author} {\bibfnamefont {Cong}\ \bibnamefont {Wang}},
  \bibinfo {author} {\bibfnamefont {Ying}\ \bibnamefont {Sun}}, \bibinfo
  {author} {\bibfnamefont {Kewen}\ \bibnamefont {Shi}}, \bibinfo {author}
  {\bibfnamefont {Sihao}\ \bibnamefont {Deng}}, \bibinfo {author}
  {\bibfnamefont {Huiqing}\ \bibnamefont {Lu}}, \bibinfo {author}
  {\bibfnamefont {Pengwei}\ \bibnamefont {Hu}}, \ and\ \bibinfo {author}
  {\bibfnamefont {Xiaoyun}\ \bibnamefont {Zhang}},\ }\bibfield  {title}
  {\enquote {\bibinfo {title} {{Metal fluorides, a new family of negative
  thermal expansion materials}},}\ }\href {\doibase 10.1016/j.jmat.2015.02.001}
  {\bibfield  {journal} {\bibinfo  {journal} {Journal of Materiomics}\ }\textbf
  {\bibinfo {volume} {1}},\ \bibinfo {pages} {106--112} (\bibinfo {year}
  {2015})}\BibitemShut {NoStop}%
\bibitem [{Note11()}]{Note11}%
  \BibitemOpen
  \bibinfo {note} {The research group of Angus Wilkinson has explored several
  fluorides with crystal structures analogous to ScF$_3$, including materials
  with various levels of doping \cite
  {Morelock:2013gi,Morelock:2014kw,Hancock:2015hl,Morelock:2015hh,Hester:2018ig}.
  Many examples display phase transitions involving rotations of octahedra and
  which appear to be continuous (second-order). These phase transitions will be
  accompanied by the softening of the RUM phonons on cooling towards the
  transition temperature, through the types of anharmonic interactions
  discussed in this paper as described by equation \ref {eq:rnpt}. We have
  shown that such a variation of the renormalised phonon frequency with
  temperature will lead to a reduction, or even elimination, of NTE \cite
  {Fang:2014cp}, as also demonstrated in recent calculations \cite
  {vanRoekeghem:2016kd,Oba:2019bi}.}\BibitemShut {Stop}%
\bibitem [{\citenamefont {Dove}\ \emph {et~al.}(2015)\citenamefont {Dove},
  \citenamefont {Tucker}, \citenamefont {Zhang}, \citenamefont {Phillips},\
  and\ \citenamefont {Du}}]{data}%
  \BibitemOpen
  \bibfield  {author} {\bibinfo {author} {\bibfnamefont {M~T}\ \bibnamefont
  {Dove}}, \bibinfo {author} {\bibfnamefont {M~G}\ \bibnamefont {Tucker}},
  \bibinfo {author} {\bibfnamefont {T}~\bibnamefont {Zhang}}, \bibinfo {author}
  {\bibfnamefont {A~E}\ \bibnamefont {Phillips}}, \ and\ \bibinfo {author}
  {\bibfnamefont {J}~\bibnamefont {Du}},\ }\href {\doibase
  10.5286/ISIS.E.RB1310476} {\enquote {\bibinfo {title} {{Structural origin of
  negative thermal expansion in ScF$_3$}},}\ } (\bibinfo {year} {2015}),\
  \bibinfo {note} {{STFC ISIS Neutron and Muon Source, DOI:
  10.5286/ISIS.E.RB1510519}}\BibitemShut {NoStop}%
\bibitem [{\citenamefont {Rimmer}\ \emph {et~al.}(2014)\citenamefont {Rimmer},
  \citenamefont {Dove}, \citenamefont {Winkler}, \citenamefont {Wilson},
  \citenamefont {Refson},\ and\ \citenamefont {Goodwin}}]{Rimmer:2014dj}%
  \BibitemOpen
  \bibfield  {author} {\bibinfo {author} {\bibfnamefont {Leila H~N}\
  \bibnamefont {Rimmer}}, \bibinfo {author} {\bibfnamefont {Martin~T}\
  \bibnamefont {Dove}}, \bibinfo {author} {\bibfnamefont {Bj{\"o}rn}\
  \bibnamefont {Winkler}}, \bibinfo {author} {\bibfnamefont {Dan~J}\
  \bibnamefont {Wilson}}, \bibinfo {author} {\bibfnamefont {Keith}\
  \bibnamefont {Refson}}, \ and\ \bibinfo {author} {\bibfnamefont {Andrew~L}\
  \bibnamefont {Goodwin}},\ }\bibfield  {title} {\enquote {\bibinfo {title}
  {{Framework flexibility and the negative thermal expansion mechanism of
  copper(I) oxide Cu$_2$O.}}}\ }\href@noop {} {\bibfield  {journal} {\bibinfo
  {journal} {Physical Review B}\ }\textbf {\bibinfo {volume} {89}},\ \bibinfo
  {pages} {214115} (\bibinfo {year} {2014})}\BibitemShut {NoStop}%
\bibitem [{\citenamefont {Tucker}\ \emph {et~al.}(2000)\citenamefont {Tucker},
  \citenamefont {Squires}, \citenamefont {Dove},\ and\ \citenamefont
  {Keen}}]{Tucker:2001vg}%
  \BibitemOpen
  \bibfield  {author} {\bibinfo {author} {\bibfnamefont {Matthew~G}\
  \bibnamefont {Tucker}}, \bibinfo {author} {\bibfnamefont {Matthew~P}\
  \bibnamefont {Squires}}, \bibinfo {author} {\bibfnamefont {Martin~T}\
  \bibnamefont {Dove}}, \ and\ \bibinfo {author} {\bibfnamefont {David~A}\
  \bibnamefont {Keen}},\ }\bibfield  {title} {\enquote {\bibinfo {title}
  {{Dynamic structural disorder in cristobalite: neutron total scattering
  measurement and reverse Monte Carlo modelling.}}}\ }\href {\doibase
  10.1088/0953-8984/13/3/304} {\bibfield  {journal} {\bibinfo  {journal}
  {Journal of Physics: Condensed Matter}\ }\textbf {\bibinfo {volume} {13}},\
  \bibinfo {pages} {403--423} (\bibinfo {year} {2000})}\BibitemShut {NoStop}%
\bibitem [{\citenamefont {Morelock}\ \emph {et~al.}(2013)\citenamefont
  {Morelock}, \citenamefont {Greve}, \citenamefont {Gallington}, \citenamefont
  {Chapman},\ and\ \citenamefont {Wilkinson}}]{Morelock:2013gi}%
  \BibitemOpen
  \bibfield  {author} {\bibinfo {author} {\bibfnamefont {Cody~R}\ \bibnamefont
  {Morelock}}, \bibinfo {author} {\bibfnamefont {Benjamin~K}\ \bibnamefont
  {Greve}}, \bibinfo {author} {\bibfnamefont {Leighanne~C}\ \bibnamefont
  {Gallington}}, \bibinfo {author} {\bibfnamefont {Karena~W}\ \bibnamefont
  {Chapman}}, \ and\ \bibinfo {author} {\bibfnamefont {Angus~P}\ \bibnamefont
  {Wilkinson}},\ }\bibfield  {title} {\enquote {\bibinfo {title} {{Negative
  thermal expansion and compressibility of Sc$_{1-x}$Y$_x$F$_3$ ($x \le
  0.25$).}}}\ }\href {\doibase 10.1063/1.4836855} {\bibfield  {journal}
  {\bibinfo  {journal} {Journal of Applied Physics}\ }\textbf {\bibinfo
  {volume} {114}},\ \bibinfo {pages} {213501} (\bibinfo {year}
  {2013})}\BibitemShut {NoStop}%
\bibitem [{\citenamefont {Morelock}\ \emph {et~al.}(2014)\citenamefont
  {Morelock}, \citenamefont {Hancock},\ and\ \citenamefont
  {Wilkinson}}]{Morelock:2014kw}%
  \BibitemOpen
  \bibfield  {author} {\bibinfo {author} {\bibfnamefont {Cody~R}\ \bibnamefont
  {Morelock}}, \bibinfo {author} {\bibfnamefont {Justin~C}\ \bibnamefont
  {Hancock}}, \ and\ \bibinfo {author} {\bibfnamefont {Angus~P}\ \bibnamefont
  {Wilkinson}},\ }\bibfield  {title} {\enquote {\bibinfo {title} {{Thermal
  expansion and phase transitions of $\alpha$-AlF$_3$}},}\ }\href {\doibase
  10.1016/j.jssc.2014.07.031} {\bibfield  {journal} {\bibinfo  {journal}
  {Journal of Solid State Chemistry}\ }\textbf {\bibinfo {volume} {219}},\
  \bibinfo {pages} {143--147} (\bibinfo {year} {2014})}\BibitemShut {NoStop}%
\bibitem [{\citenamefont {Hancock}\ \emph {et~al.}(2015)\citenamefont
  {Hancock}, \citenamefont {Chapman}, \citenamefont {Halder}, \citenamefont
  {Morelock}, \citenamefont {Kaplan}, \citenamefont {Gallington}, \citenamefont
  {Bongiorno}, \citenamefont {Han}, \citenamefont {Zhou},\ and\ \citenamefont
  {Wilkinson}}]{Hancock:2015hl}%
  \BibitemOpen
  \bibfield  {author} {\bibinfo {author} {\bibfnamefont {Justin~C}\
  \bibnamefont {Hancock}}, \bibinfo {author} {\bibfnamefont {Karena~W}\
  \bibnamefont {Chapman}}, \bibinfo {author} {\bibfnamefont {Gregory~J}\
  \bibnamefont {Halder}}, \bibinfo {author} {\bibfnamefont {Cody~R}\
  \bibnamefont {Morelock}}, \bibinfo {author} {\bibfnamefont {Benjamin~S}\
  \bibnamefont {Kaplan}}, \bibinfo {author} {\bibfnamefont {Leighanne~C}\
  \bibnamefont {Gallington}}, \bibinfo {author} {\bibfnamefont {Angelo}\
  \bibnamefont {Bongiorno}}, \bibinfo {author} {\bibfnamefont {Chu}\
  \bibnamefont {Han}}, \bibinfo {author} {\bibfnamefont {Si}~\bibnamefont
  {Zhou}}, \ and\ \bibinfo {author} {\bibfnamefont {Angus~P}\ \bibnamefont
  {Wilkinson}},\ }\bibfield  {title} {\enquote {\bibinfo {title} {{Large
  negative thermal expansion and anomalous behavior on compression in cubic
  ReO$_3$-type A$^\mathrm{II}$B$^\mathrm{IV}$F$_6$: CaZrF$_6$ and
  CaHfF$_6$.}}}\ }\href {\doibase 10.1021/acs.chemmater.5b00662} {\bibfield
  {journal} {\bibinfo  {journal} {Chemistry of Materials}\ }\textbf {\bibinfo
  {volume} {27}},\ \bibinfo {pages} {3912--3918} (\bibinfo {year}
  {2015})}\BibitemShut {NoStop}%
\bibitem [{\citenamefont {Morelock}\ \emph {et~al.}(2015)\citenamefont
  {Morelock}, \citenamefont {Gallington},\ and\ \citenamefont
  {Wilkinson}}]{Morelock:2015hh}%
  \BibitemOpen
  \bibfield  {author} {\bibinfo {author} {\bibfnamefont {Cody~R}\ \bibnamefont
  {Morelock}}, \bibinfo {author} {\bibfnamefont {Leighanne~C}\ \bibnamefont
  {Gallington}}, \ and\ \bibinfo {author} {\bibfnamefont {Angus~P}\
  \bibnamefont {Wilkinson}},\ }\bibfield  {title} {\enquote {\bibinfo {title}
  {{Solid solubility, phase transitions, thermal expansion, and compressibility
  in Sc$_{1-x}$Al$_x$F$_3$.}}}\ }\href {\doibase 10.1016/j.jssc.2014.11.007}
  {\bibfield  {journal} {\bibinfo  {journal} {Journal of Solid State
  Chemistry}\ }\textbf {\bibinfo {volume} {222}},\ \bibinfo {pages} {96--102}
  (\bibinfo {year} {2015})}\BibitemShut {NoStop}%
\bibitem [{\citenamefont {Hester}\ and\ \citenamefont
  {Wilkinson}(2018)}]{Hester:2018ig}%
  \BibitemOpen
  \bibfield  {author} {\bibinfo {author} {\bibfnamefont {Brett~R}\ \bibnamefont
  {Hester}}\ and\ \bibinfo {author} {\bibfnamefont {Angus~P}\ \bibnamefont
  {Wilkinson}},\ }\bibfield  {title} {\enquote {\bibinfo {title} {{Negative
  Thermal Expansion, Response to Pressure and Phase Transitions in
  CaTiF$_6$}},}\ }\href {\doibase 10.1021/acs.inorgchem.8b01912} {\bibfield
  {journal} {\bibinfo  {journal} {Inorganic Chemistry}\ }\textbf {\bibinfo
  {volume} {57}},\ \bibinfo {pages} {11275--11281} (\bibinfo {year}
  {2018})}\BibitemShut {NoStop}%
\end{thebibliography}


\end{document}